\documentclass[twoside,twocolumn,9pt]{article}
\usepackage{extsizes}
\usepackage[super,sort&compress,comma]{natbib} 
\usepackage[version=3]{mhchem}
\usepackage[left=1.5cm, right=1.5cm, top=1.785cm, bottom=2.0cm]{geometry}
\usepackage{balance}
\usepackage{times,mathptmx}
\usepackage{sectsty}
\usepackage{graphicx} 
\usepackage{lastpage}
\usepackage[format=plain,justification=justified,singlelinecheck=false,font={stretch=1.125,small,sf},labelfont=bf,labelsep=space]{caption}
\usepackage{float}
\usepackage{fancyhdr}
\usepackage{fnpos}
\usepackage[english]{babel}
\usepackage{array}
\usepackage{droidsans}
\usepackage{charter}
\usepackage[T1]{fontenc}
\usepackage[usenames,dvipsnames]{xcolor}
\usepackage{setspace}
\usepackage[compact]{titlesec}
\usepackage{hyperref}
\usepackage[counterclockwise,figuresright]{rotating}
\usepackage{dcolumn}

\usepackage{booktabs} %beautiful tables
\usepackage{textcomp} % for straight mu (among other things)

\usepackage{epstopdf}%This line makes .eps figures into .pdf - please comment out if not required.

\definecolor{cream}{RGB}{222,217,201}

\begin{document}

\pagestyle{fancy}
\thispagestyle{plain}
\fancypagestyle{plain}{

%%%HEADER%%%
\fancyhead[C]{\includegraphics[width=18.5cm]{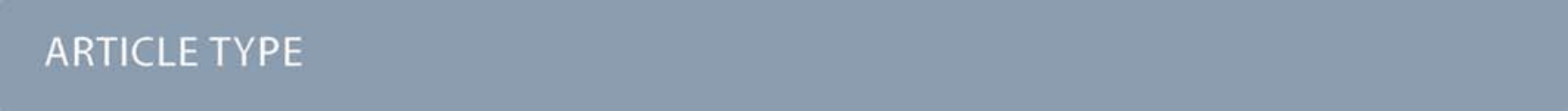}}
\fancyhead[L]{\hspace{0cm}\vspace{1.5cm}\includegraphics[height=30pt]{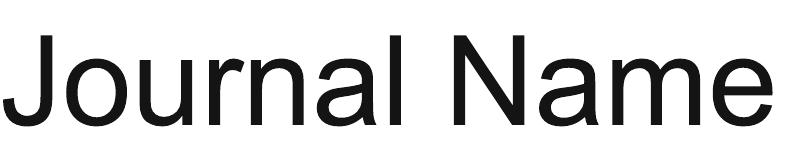}}
\fancyhead[R]{\hspace{0cm}\vspace{1.7cm}\includegraphics[height=55pt]{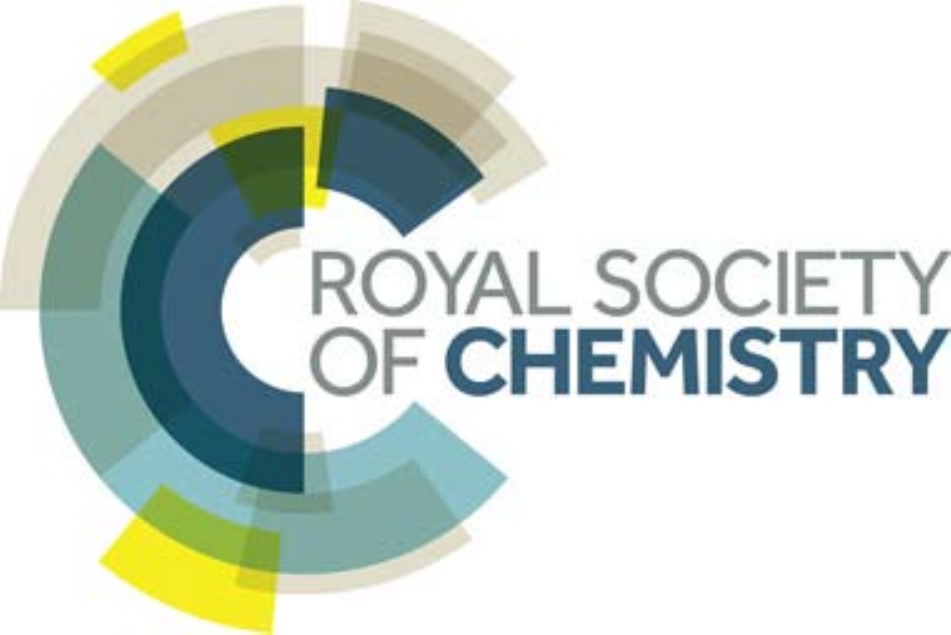}}
\renewcommand{\headrulewidth}{0pt}
}
%%%END OF HEADER%%%

%%%PAGE SETUP - Please do not change any commands within this section%%%
\makeFNbottom
\makeatletter
\renewcommand\LARGE{\@setfontsize\LARGE{15pt}{17}}
\renewcommand\Large{\@setfontsize\Large{12pt}{14}}
\renewcommand\large{\@setfontsize\large{10pt}{12}}
\renewcommand\footnotesize{\@setfontsize\footnotesize{7pt}{10}}
\makeatother

\renewcommand{\thefootnote}{\fnsymbol{footnote}}
\renewcommand\footnoterule{\vspace*{1pt}% 
\color{cream}\hrule width 3.5in height 0.4pt \color{black}\vspace*{5pt}} 
\setcounter{secnumdepth}{5}

\makeatletter 
\renewcommand\@biblabel[1]{#1}            
\renewcommand\@makefntext[1]% 
{\noindent\makebox[0pt][r]{\@thefnmark\,}#1}
\makeatother 
\renewcommand{\figurename}{\small{Fig.}~}
\sectionfont{\sffamily\Large}
\subsectionfont{\normalsize}
\subsubsectionfont{\bf}
\setstretch{1.125} %In particular, please do not alter this line.
\setlength{\skip\footins}{0.8cm}
\setlength{\footnotesep}{0.25cm}
\setlength{\jot}{10pt}
\titlespacing*{\section}{0pt}{4pt}{4pt}
\titlespacing*{\subsection}{0pt}{15pt}{1pt}
%%%END OF PAGE SETUP%%%

%%%FOOTER%%%
\fancyfoot{}
\fancyfoot[LO,RE]{\vspace{-7.1pt}\includegraphics[height=9pt]{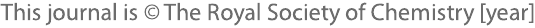}}
\fancyfoot[CO]{\vspace{-7.1pt}\hspace{13.2cm}\includegraphics{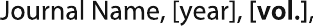}}
\fancyfoot[CE]{\vspace{-7.2pt}\hspace{-14.2cm}\includegraphics{head_foot/RF}}
\fancyfoot[RO]{\footnotesize{\sffamily{1--\pageref{LastPage} ~\textbar  \hspace{2pt}\thepage}}}
\fancyfoot[LE]{\footnotesize{\sffamily{\thepage~\textbar\hspace{3.45cm} 1--\pageref{LastPage}}}}
\fancyhead{}
\renewcommand{\headrulewidth}{0pt} 
\renewcommand{\footrulewidth}{0pt}
\setlength{\arrayrulewidth}{1pt}
\setlength{\columnsep}{6.5mm}
\setlength\bibsep{1pt}
%%%END OF FOOTER%%%

%%%FIGURE SETUP - please do not change any commands within this section%%%
\makeatletter 
\newlength{\figrulesep} 
\setlength{\figrulesep}{0.5\textfloatsep} 

\newcommand{\topfigrule}{\vspace*{-1pt}% 
\noindent{\color{cream}\rule[-\figrulesep]{\columnwidth}{1.5pt}} }

\newcommand{\botfigrule}{\vspace*{-2pt}% 
\noindent{\color{cream}\rule[\figrulesep]{\columnwidth}{1.5pt}} }

\newcommand{\dblfigrule}{\vspace*{-1pt}% 
\noindent{\color{cream}\rule[-\figrulesep]{\textwidth}{1.5pt}} }

\makeatother
%%%END OF FIGURE SETUP%%%

%%%TITLE, AUTHORS AND ABSTRACT%%%
\twocolumn[
  \begin{@twocolumnfalse}
\vspace{3cm}
\sffamily
\begin{tabular}{m{4.5cm} p{13.5cm} }

\includegraphics{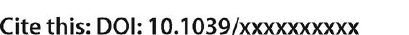} & \noindent\LARGE{\textbf{Vibrational Satellites of \ce{C2S}, \ce{C3S}, and \ce{C4S}: \hspace{5in} Microwave Spectral Taxonomy as a Stepping Stone to the Millimeter-Wave Band$^\dag$}} \\
\vspace{0.3cm} & \vspace{0.3cm} \\

 & \noindent\large{Brett A. McGuire,$^{a,b,\ddag}$ Marie-Aline Martin-Drumel,$^{b, \S}$ Kin Long Kelvin Lee,$^b$  John F. Stanton,$^c$ Carl A. Gottlieb,$^b$ and Michael C. McCarthy$^{\ast,b,d}$} \\

\includegraphics{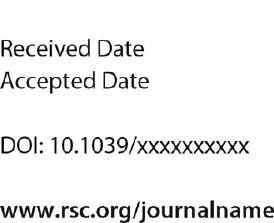} & \noindent\normalsize{We present a microwave spectral taxonomy study of several hydrocarbon/\ce{CS2} discharge mixtures in which more than 60 distinct chemical species, their more abundant isotopic species, and/or their vibrationally excited states  were detected using chirped-pulse and cavity Fourier-transform microwave spectroscopies. Taken together, in excess of 85 unique variants were detected, including several new isotopic species and more than 25 new vibrationally excited states of \ce{C2S}, \ce{C3S}, and  \ce{C4S}, which have been assigned on the basis of published vibration-rotation interaction constants for \ce{C3S}, or newly calculated ones for \ce{C2S} and \ce{C4S}.  On the basis of these precise, low-frequency measurements, several vibrationally exited states of \ce{C2S} and \ce{C3S} were subsequently identified in archival millimeter-wave data in the 253--280~GHz frequency range, ultimately providing highly accurate catalogs for astronomical searches.  As part of this work, formation pathways of the two smaller carbon-sulfur chains were investigated using $^{13}$C isotopic spectroscopy, as was their vibrational excitation.  The present study illustrates the utility of microwave spectral taxonomy as a tool for complex mixture analysis, and as a powerful and convenient `stepping stone' to higher frequency measurements in the millimeter and submillimeter bands.}\\

\end{tabular}

 \end{@twocolumnfalse} \vspace{0.6cm}

  ]
%%%END OF TITLE, AUTHORS AND ABSTRACT%%%

%%%FONT SETUP - please do not change any commands within this section
\renewcommand*\rmdefault{bch}\normalfont\upshape
\rmfamily
\section*{}
\vspace{-1cm}

%%%FOOTNOTES%%%

\footnotetext{$^a$~National Radio Astronomy Observatory, 520 Edgemont Rd, Charlottesville, VA USA 22903.\\
$^b$~Harvard-Smithsonian Center for Astrophysics, 60 Garden Street, Cambridge, MA USA 02138.\\
$^c$~Department of Chemistry, University of Florida, Gainesville, FL USA 32611.\\
$^d$~School of Engineering and Applied Sciences, Harvard University, 29 Oxford Street, Cambridge, MA USA 02138.\\
$^{\ddag}$~B.A.M. is a Hubble Fellow of the National Radio Astronomy Observatory.\\
$^{\S}$~Present address: Institut des Sciences Mol{\'e}culaires d'Orsay, CNRS, Universit{\'e} Paris-Sud, Univ. Paris-Saclay, Orsay, France\\
}

%Please use \dag to cite the ESI in the main text of the article.
%If you article does not have ESI please remove the the \dag symbol from the title and the footnotetext below.
\footnotetext{\dag~Electronic Supplementary Information (ESI) available: [details of any supplementary information available should be included here]. See DOI: 10.1039/b000000x/}
%additional addresses can be cited as above using the lower-case letters, c, d, e... If all authors are from the same address, no letter is required

%%%END OF FOOTNOTES%%%

%%%MAIN TEXT%%%%

\section{Introduction}
Quantitative chemical analysis of complex mixtures is of interest to a broad range of fields ranging from atmospheric and combustion science,\cite{Taatjes:2008cl} to the food industry,\cite{Semmelroch:1995hy} and astrochemistry.\cite{Loomis:2013fs} Because of their high sensitivity, mass spectrometry and gas chromatography, either separately or in combination, are widely used analytical techniques.  Although capable of discriminating mixtures comprising ${\sim}$100 compounds~\cite{kowaklick:230,teixeira:9875}, both techniques become laborious as the number of components increases, and may lack unambiguous molecular specificity for large compounds.  Recent studies of flames of 2,5-dimethylfuran -- a promising biofuel alternative to ethanol due to its higher energy density \cite{Daniel:2011gl} and ease of production from biological sources -- provide a good illustration of the strengths and weaknesses of these approaches.\cite{Rosatella:2011ju}  In the study by Wu \textit{et al.}, the combustion products of 2,5-dimethylfuran were investigated using molecular beam photoionization mass spectrometry (PIMS). \cite{Wu:2009ji} The same system was studied again in 2014 by a separate team using gas chromatography.\cite{TogbA:2014hm} Although the two studies agree on many compounds, there are marked differences in the molecular assignments of the C$_2$H$_6$O, C$_4$H$_6$O, C$_4$H$_8$O, and C$_5$H$_8$O isomers.  For example, where one study assigned signal from C$_2$H$_6$O to be dimethyl ether (CH$_3$OCH$_3$),\cite{TogbA:2014hm} the other does not.

Advances in microwave spectroscopy in the last decade provide a promising, complementary approach to complex mixture analysis.  The development of broadband or chirped-pulse (CP) Fourier transform microwave (FTMW) spectroscopy has revolutionized the field, allowing data over many GHz of bandwidth to be collected simultaneously.\cite{Brown:2008gk} As the spectral resolution is normally very high, and rotational transitions provide a unique diagnostic for each chemical species since they are dictated by the three moments of inertia of the species, it is now possible to identify the presence of many chemical compounds in a mixture and quantify their abundance with no ambiguity in the atomic connectivity or molecular structure.   Because a rich array of astronomical molecules can often be produced when an electrical discharge is combined with a supersonic jet source \cite{Crabtree:2016fj}, much laboratory effort has been devoted to characterizing the resulting rotational spectra, with the expectation that entirely new species of astronomical interest might be detected. Due to the non-specificity of this production method, however, the simultaneous production of both familiar and exotic molecules creates a challenge in rapid spectral analysis and identification.

%Here, we utilize a new CP-FTMW technique, Microwave Spectral Taxonomy (MST) in the context of screening astrochemically-relevant reaction mixtures.\cite{Crabtree:2016fj}

Astronomical sources, analogous to biofuels, are extremely complex mixtures due to their highly diverse and unusual chemistry: conditions depart significantly from thermodynamic equilibrium,\cite{Garrod:2008tk} and are known to have considerable spatio-chemical variation.  This complexity has become even more apparent and daunting with the advent of powerful radio interferometers, specifically the Atacama Large Millimeter/sub-millimeter Array (ALMA), which can perform spectral line surveys in 8\,GHz frequency intervals with unprecedented sensitivity and angular resolution.  In doing so, many new spectral lines have been reported, but a sizable fraction of these remain unassigned due to the absence of supporting laboratory data.\cite{Belloche:2016fm,Jorgensen:2016cq,Cernicharo:2013cc} Some of these unassigned astronomical lines may arise from entirely new molecules, which are critical to advancing our understanding of interstellar chemistry. Equally likely, however, is that many instead arise from a relatively small number of highly abundant, known interstellar species, but in previously unanalyzed, low-lying vibrational states or isotopic forms.\cite{Fortman:2012is} 

One of the richest and most chemically diverse astronomical sources is IRC+10216, a carbon-rich evolved star.  Nearly 50\% of the nearly 200 known astronomical molecules have been observed there, including unsaturated carbon and sulfur-containing species in vibrationally excited states.\cite{Cernicharo:2013cc,Cernicharo:2010es,Cernicharo:2008wi}   While other chemically rich sources such as Sgr B2(N) are challenging to analyze due to the complexity brought about by line-confusion, especially in the ALMA era\cite{Belloche:2016fm}, spectra of IRC+10216 do not yet approach this limit, even at high sensitivity.  Nevertheless, the number of  unassigned features is shockingly large.\cite{Cernicharo:2013cc}  For this reason there is great value in conducting laboratory investigations that mimic --- in a very general sense --- the chemistry in IRC+10216, in an attempt to understand the rich but enigmatic spectrum of this source.

The traditional experimental approach to investigating the rotational spectrum of a molecule is a successful, if laborious, procedure in which the species of interest is selectively produced, and quantum numbers are then assigned to new transitions on the basis of a model Hamiltonian in a largely step-wise fashion.\cite{Muller:2016de}  This method has the benefit of providing predictive fits which are nominally accurate for lines not directly measured in the laboratory; the extent to which this extrapolation remains valid is closely related to the spectral complexity of the molecule and the robustness of the model.\cite{Carroll:2010gt}

Another approach, pioneered in recent years, eschews the assignment of quantum numbers, and instead provides a `complete' list of frequencies and intensities for all transitions of a single molecule observed within a narrowly-constrained range of excitation temperatures.\cite{Medvedev:2007kl}  This approach has been used to successfully identify a significant number of previously unassigned lines in molecular line survey data, \cite{Fortman:2012is} and has the distinct advantage of not needing to selectively target one vibrational state at a time for analysis.  All excited states which have a detectable population at a given excitation temperature are analyzed and cataloged. While it has the merit of including large numbers of transitions which may be missing from the traditional Hamiltonian-based catalogs, there are two main drawbacks: first, the end-product line catalogs have no predictive power, and therefore cannot be used outside the frequency -- and to a lesser extent temperature -- range of the experiment, and second, the non-specificity in target selection is itself problematic: in astronomical spectra, there is no obvious way to readily distinguish between weak lines of the ground vibrational state and lines of excited vibrational states.  We recently developed an alternative approach, microwave spectral taxonomy (MST), to identify unknown species in complex mixtures -- new molecules as well as vibrational satellites and isotopologues of known species -- without an \textit{a priori} bias of atomic composition or molecular structure.\cite{Crabtree:2016fj} 

Here, we present a MST reaction-screening study of C- and S-chemistry relevant to an astronomical source such as IRC+10216.  By using a S-bearing precursor gas, carbon disulfide (\ce{CS2}), and either acetylene (HCCH) or diacetylene (\ce{HC4H}), and subjecting these reactants to a dc electric discharge in combination with a supersonic jet, a rich array of transient species, many of direct relevance to the chemistry of IRC+10216, were produced in high abundance.  In the course of our analysis, many entirely new spectral lines were observed with high signal-to-noise ratios (SNRs), and found to be lie close in frequency to those predicted from published or new theoretical vibration-rotation interaction constants for \ce{C2S}, \ce{C3S}, or \ce{C4S}. By extrapolating these precise low-frequency measurements to high $J$, attempts were then made to detect higher frequency transitions in legacy millimeter-wavelength direct-absorption spectra starting from the same reactants.   The large number of new states enables one to study mode-specific excitation in each chain in detail and the extent of vibrational excitation as a function of chain length; $^{13}$C isotopic studies have also been performed to test possible formation pathways for  \ce{C2S} and \ce{C3S} in our discharge source.

%%%%%%%%%%%%%%%%%%%%%%%%%%%%%%%%%%%%%%%%%%%%%%%%%%%%%%%%%%%%%%%%%%%%%%%%%%%%%%%%%%%%%%%%%
%%%%%%%%%%%%%%%%%%%%%%%%%%%  methodology               %%%%%%%%%%%%%%%%%%%%%%%%%%%%%%%%%%
%%%%%%%%%%%%%%%%%%%%%%%%%%%%%%%%%%%%%%%%%%%%%%%%%%%%%%%%%%%%%%%%%%%%%%%%%%%%%%%%%%%%%%%%%

\section{Methodology}

\subsection{Spectroscopy of \ce{C2S}, \ce{C3S}, and \ce{C4S}}

\begin{figure*}[ht!]
\includegraphics[width=1.\textwidth]{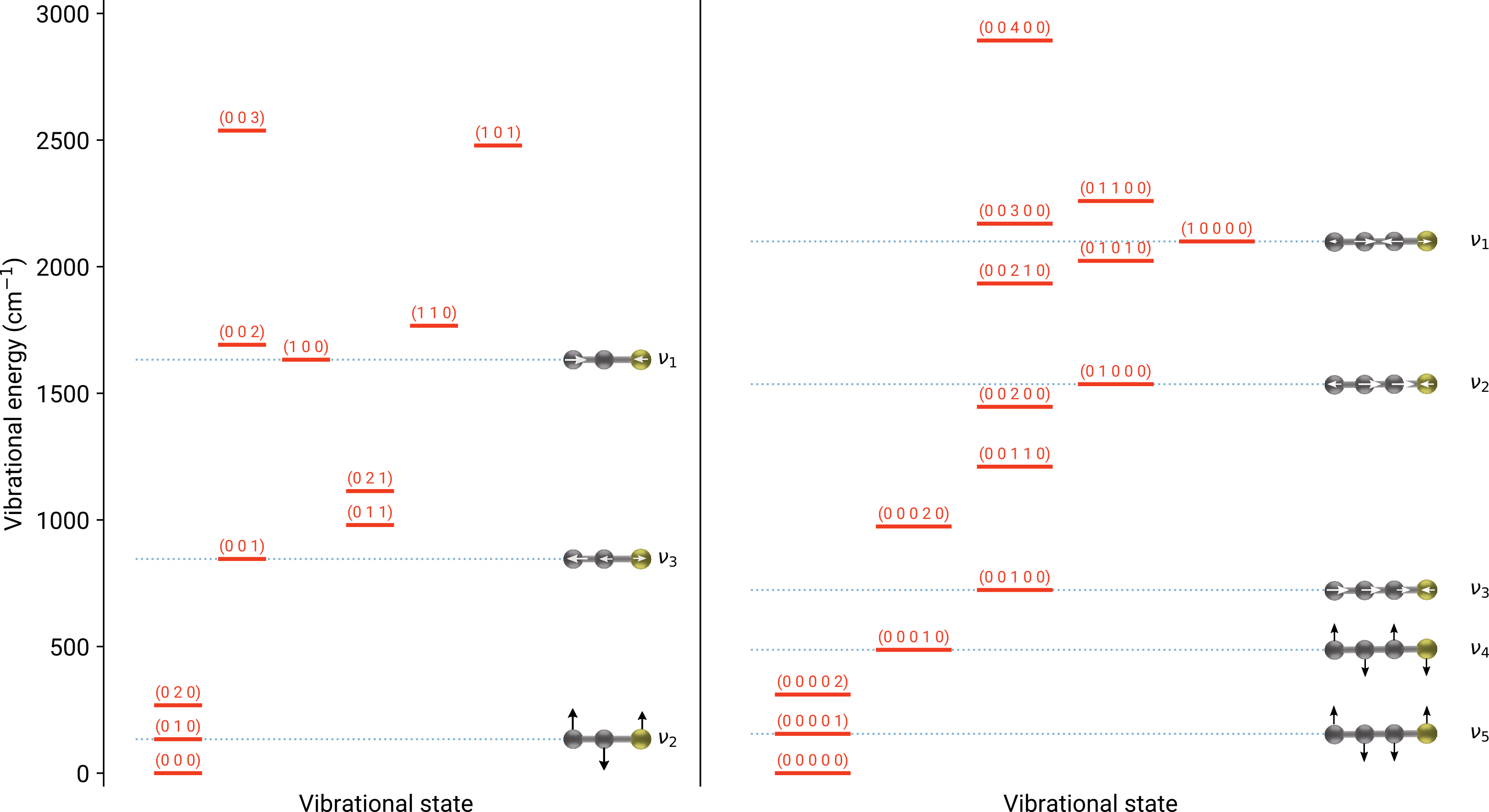}
    \caption{Vibrational energy level diagram of \ce{C2S} (left) and \ce{C3S} (right). The mode deformation associated with each fundamental vibration is represented pictorially. Vibrational quantum numbers for each state are indicated using the convention ($v_1\ v_2\ v_3 [\cdots]$) in which $v_i$ is the quanta of excitation in the $\nu_i$ modes. For brevity, the $l$ quantum number associated with the bending modes has been omitted. For \ce{C2S}, levels for which pure rotational transitions have been observed in this work are shown with plain lines (in red), while some indicative levels which are not observed are represented with dashed lines (in gray).  Owing to the high density of states for \ce{C3S}, only observed and fundamental vibrational states are shown.}
    \label{egy_diag}
\end{figure*}

Both \ce{C2S} and \ce{C4S} are open-shell molecules possessing $^3\Sigma^-$ electronic ground states \cite{Saito:1987fa, Hirahara:1993ud}. \ce{C2S} possesses three vibrational modes, two stretching, $\nu_1$ at 1634\,cm$^{-1}$ and $\nu_3$ at 846\,cm$^{-1}$ above ground, and a doubly-degenerate bending $\nu_2$ at 134\,cm$^{-1}$ (Table~\ref{abinitio_data}); \ce{C4S} possesses a total of seven modes, four stretching and three bending modes (Table~\ref{abinitio_data}). \ce{C3S} has a closed-shell $^1\Sigma^+$ electronic ground state\cite{Yamamoto:1987jd} and five fundamental vibrations, three stretches, $\nu_1$ to $\nu_3$, respectively at 2046, 1560, 731 cm$^{-1}$, and two doubly-degenerated bends, $\nu_4$ and $\nu_5$ at 490 and 150 cm$^{-1}$. Fig. \ref{egy_diag} and \ref{egy_diag_c4s} show the vibrational energy level diagram of \ce{C2S}, \ce{C3S}, and \ce{C4S} along with the deformation associated with each vibration.

As with any linear polyatomic molecule, vibrationally excited states with one or more quanta of excitation in a bending mode require an additional quantum number, $l$, with $l=\sum\limits_i l_i$ where $l_i=|v_i|, |v_i-2|, |v_i-4|, \cdots$ with $v_i$ the quanta of excitation in the $\nu_i$ mode. Selection rules for pure rotational transitions are $\Delta J = \pm 1$, $\Delta l = 0$. As a consequence, pure rotational transitions from the first excited state of a bending $l=\pm 1$ mode will appear as doublets, while triplets are expected for the second excited state ($l=0, \pm 2$), etc.

To avoid confusion or ambiguity in notation, we have adopted the following convention throughout the paper: vibrational states are labelled simply as $v_i=u_i$, or when appropriate $v_i=u_i^{l_i}$ and $(v_i,v_j)=(u_i, u_j)$, if more than one vibration is excited with $u_i/ u_j$ the quanta of excitation in the 
$\nu_i/ \nu_j$ modes, respectively, e.g., $v_2=1$ or $(v_3, v_4)= (2, 2^0)$. The only exception is Figure~\ref{egy_diag}, where the notation ($v_1\ v_2\ v_3$) for \ce{C2S} and ($v_1\ v_2\ v_3\ v_4\ v_5$) for \ce{C3S} has been used for the sake of simplicity (Fig. \ref{egy_diag_c4s} adopts the same formalism for \ce{C4S}).

\subsection{Microwave spectral taxonomy}

In the centimeter-wave regime, FTMW techniques have been a workhorse of molecular spectroscopy for more than 30 years.\cite{Balle:1981ex}  In either the CP or cavity variants of this method, a short pulse of microwave radiation, typically of order a few $\mu$s, creates a coherent macroscopic polarization, provided that one or molecules possesses a rotational transition in the FT-limited frequency bandwidth of the pulse.   Because the timescale of the free-induction decay (FID) is typically at least an order of magnitude longer than the excitation step, the molecular signal can be detected with a sensitive microwave receiver in a `background-free' regime.  The Fourier-transform of the time-domain signal in combination with the frequency of the microwave radiation yields the precise frequencies of the rotational transitions.  

Cavity FTMW spectroscopy is widely used because it provides high sensitivity, albeit with a very narrow instantaneous bandwidth, of order 0.5\,MHz at each setting of the Fabry-Perot cavity. In contrast, CP-FTMW extends the instantaneous spectral coverage to many GHz --- at a modest reduction (a factor of $\sim$40) in sensitivity and spectral resolution (a factor of $\sim$10) compared to the cavity variant --- allowing the acquisition of enormous portions of a rotational spectrum simultaneously.\cite{Brown:2008gk}  Because CP-FTMW has a fairly flat instrumental response, relative abundances of multiple species can often be determined with far greater accuracy than with cavity measurements.

To exploit the unique strengths of CP-FTMW and cavity-FTMW, MST was recently developed. In this procedure, the chirped spectrum is first used as a survey tool to detect active, or `bright,' resolution elements. The spectral features lying within the frequency range of the CP-FTMW spectrum (typically 2--8 or 8--18\,GHz) are analyzed using a program such as \textsc{SPECData}, which is a newly developed in-house database query system that rapidly assigns transitions of known species, common contaminants, and instrumental artifacts, in a semi-automated fashion.\cite{jasmine:isms} The remaining, unidentified features are then scrutinized using cavity spectroscopy for rapid characterization, because of its higher instantaneous sensitivity and spectral resolution. Part of this analysis is to categorize each spectral line based on a set of quantifiable properties: dipole moment; precursor dependence; magnet susceptibility; requirement of an excitation source (e.g., electrical discharge or laser ablation). Following classification, lines that share a common set of characteristics may be then subjected to exhaustive double resonance (DR) tests to determine those transitions that share a common upper or lower energy level, and thus arise from the same carrier.  Previous studies of large aromatic compounds\cite{MartinDrumel:2016fw} found that only a handful of linkages may be needed to determine all three rotational constants of a molecule.  Details of the two spectrometers used in the present investigation have been described in detail elsewhere\cite{Crabtree:2016fj}, and will only be discussed briefly.

\subsection{Centimeter-wave Measurements}

In this study, CP-FTMW spectra were acquired in the 7.5--18\,GHz band using several precursor combinations: a sulfur source (\ce{CS2}) and one of two hydrocarbons (HCCH and \ce{HC4H}), heavily diluted in argon, and then expanded in neon. The various gas mixtures, their concentrations, and total number of FIDs acquired in each experiment are summarized in Table~\ref{ratios}.  In all cases, the gas mixture was subject to a dc discharge just after the pulsed nozzle source, but prior to supersonic expansion into the large vacuum chamber. The pulsed nozzle was operated at a repetition rate of 5\,kHz, the backing pressure behind the nozzle was 2.5\,kTorr (333\,kPa), and the discharge voltage was typically 1.5\,kV.  Finally, 10 FIDs were collected per gas pulse during the CP-FTMW measurements. In addition to recording spectra of the two \ce{CS2}/hydrocarbon mixtures, spectra starting with only \ce{CS2} or \ce{HC4H} were also acquired, to determine which species required the presence of both reactants.  

After acquisition, features with a specified SNR (greater than five) were automatically flagged in each CP-FTMW spectrum, and subjected to further processing to identify those that coincided with transitions of well-known molecules or are instrumental in origin.  Roughly 50\% of the original features were removed in this step; the remaining features were then scrutinized further with a  cavity FTMW spectrometer using essentially identical experimental conditions.  A by-product of the cavity studies is improved frequency accuracy (2\,kHz or better) compared with CP-FTMW spectroscopy, by roughly an order of magnitude (10 to 50). If a new series of nearly harmonically related lines was identified between 7.5 and 18\,GHz, the measurements were routinely extended to 40\,GHz in the cavity instrument with the same level of accuracy. Double resonance techniques were also used to extend measurements above the frequency range of the cavity instrument, providing that the higher-frequency line shared an energy level with a strong centimeter-wave line (see Tables \ref{c2s_cm_freqs} and \ref{c2s_mm_freqs}).

\begin{table}[ht]
\small
  \caption{Precursor gases (diluted in Ar), mixing ratios, and number of averages collected in the CP-FTMW experiments.}
  \label{ratios}
  \begin{tabular*}{0.5\textwidth}{@{\extracolsep{\fill}}l c c r}
    \toprule
   Precursor Gases & Mixing Ratio$^a$ & dc voltage /kV& FIDs /million\\
    \midrule
    2\% \ce{CS2} : 0.75\% \ce{HC4H} &	1:1:7	& 1.15 &	1.25	\\
    1\% \ce{CS2} : 2\% \ce{HCCH} 	&	3:6:5	& 1    &	1.60	\\ 
    2\% \ce{CS2} 					&	1:9		& 1.15  &	1.25 	\\   
    0.75\% \ce{HC4H} 				&	1:4		& 1.15  &	1.25  	\\
    \bottomrule
  \end{tabular*}
  
    \smallskip
    \begin{minipage}{0.5\textwidth}
        $^a$ The buffer gas was Ne in each experiment and its ratio is the last value reported.
    \end{minipage}
\end{table}

\subsection{Millimeter-wave Measurements\label{mmw}}

Evidence for millimeter-wave lines of vibrationally excited \ce{C2S} and \ce{C3S} was found in archival spectra taken during previous searches for the HCCS \cite{Vrtilek:1992jt} and \ce{HC3S} radicals\cite{mccarthy:l127}.  Spectra were acquired between 253 and 280\,GHz using a 3\,m long free space absorption spectrometer that has been described in detail previously~\cite{mccarthy:7779,gottlieb:655}, in which a low pressure (35\,mTorr) dc discharge (200\,mA) was struck through \ce{HCCH}, \ce{CS2}, and helium (He) in a molar ratio of 10:5:1, with the walls of the discharge cell cooled to 190\,K using liquid nitrogen.  In this frequency range, strong lines of both \ce{HCCS} and \ce{HC3S} were observed, as were lines of \ce{C2S} and \ce{C3S}.  For  \ce{C2S}, two rotational transitions ($J'=21 - 20$ and $20 - 19$) were covered, while five transitions ($J'=48 - 44$) of \ce{C3S} lie in the same frequency range due to its smaller rotational constant. 

Tunable millimeter-wave radiation, generated by a phase-locked Gunn oscillator in combination with a frequency multiplier ($\times$2 or $\times$3), passed twice through the absorption cell to improve the SNR. Before entering the cell, radiation first passed through a grid polarizer and then, after passing through a lens, propagated along the length of the discharge cell where it was then reflected by a roof top mirror; reflection rotates the plane of polarization by 90$^{\circ}$. After counter-propagating back through the cell, radiation was reflected by the grid polarizer and focused onto a sensitive, liquid He cooled indium antimonide (InSb) detector. To suppress 1/$f$ noise, frequency modulation combined with lock-in detection at $2f$ was used, resulting in line profiles that are well described by the second-derivative of a Lorentzian.  

Because the Gunn oscillator is a resonant device with limited frequency agility, spectra were acquired in 200\,MHz segments before the  oscillator required manual re-tuning; the resulting segments were then concatenated together to produce a survey with continuous frequency coverage over many GHz. To distinguish between rotational lines of radicals and non-radicals, each frequency segment was recorded twice, once in the absence of a strong axial magnetic field, and then again in the presence of the magnetic field using otherwise identical conditions.  By subtracting these two spectra, it is possible to identify only open-shell species such as \ce{C2S}, as non-magnetic lines are typically subtracted out, with residuals at the level of a few percent. In contrast, lines of closed-shell \ce{C3S} are expected to be present in both spectra.  Because each rotational transition of \ce{C2S} is magnetic and consists of a closely spaced triplet, this combination provided a distinct spectral signature for new lines of \ce{C2S}, even though only two of its transitions fall in the range of the existing survey.

\subsection{Quantum chemical calculations\label{qcc}}

Calculations of the molecular structures and vibration-rotation interactions were performed using the CFOUR suite of electronic structure programs. \cite{stanton_j_f_cfour_2017} The molecular geometries of \ce{C2S} and \ce{C4S} were optimized using coupled-cluster methods with single, double, and perturbative triple excitations [CCSD(T)], based on an unrestricted Hartree-Fock (UHF) reference wavefunction to treat the triplet multiplicity of these species. The calculations were performed with the frozen-core (fc) approximation using the correlation consistent basis sets of Dunning (i.e. cc-pVXZ). \cite{DunningBasis} The geometry optimizations were converged to a root-mean-squared value for the molecular gradient to less than $10^{-7}$ hartrees/bohr. The resulting structures were verified to be minimum energy geometries by harmonic frequency analysis. Subsequently, the vibration-rotation coupling constants were calculated to first order ($\alpha_i$) using second-order vibrational perturbation theory (VPT2) as implemented in CFOUR, with the required cubic force-fields computed via finite-differences of analytic gradients. 

The exothermocity of the reaction between linear \ce{C3} and \ce{S} ($\mathrm{^3P}$) were calculated using the HEAT345(Q) scheme. The method is well-documented in previous publications,\cite{harding_high-accuracy_2008,bomble_high-accuracy_2006} and thus only briefly summarized here. The molecular geometries of linear \ce{C3} and \ce{C3S} are first optimized at the ae-CCSD(T)/cc-pVQZ level of theory. Based on the structure obtained at this level, the HEAT345(Q) energy ($E_\mathrm{HEAT}$) is given by the sum of additive terms:
\begin{equation}
E_\mathrm{HEAT} = E^\infty _\mathrm{HF} + E^\infty_\mathrm{CCSD(T)} + E_{\Delta \mathrm{T - (T)}} + E_{\mathrm{(Q)}} + E_\mathrm{ZPE} + E_\mathrm{DBOC} + E_\mathrm{Rel.}
\label{eq:heat}
\end{equation}
where $E^\infty _\mathrm{HF}$ and $E^\infty_\mathrm{CCSD(T)}$ are the extrapolated Hartree-Fock and CCSD(T) correlation contributions based on calculations with aug-cc-pCVXZ (where X=T,Q,5) basis, $E_{\Delta \mathrm{T - (T)}}$ is the extrapolated difference between the fc-CCSDT and fc-CCSD(T) energies with cc-pVXZ (where X=T,Q) basis, $E_{\Delta\mathrm{HLC}}$, $E_{(Q)}$ is the correlation contribution from perturbative quadruple excitations from an fc-CCSDT(Q)/cc-pVDZ calculation, $E_\mathrm{ZPE}$ is the harmonic zero-point energy, $E_\mathrm{DBOC}$ is the diagonal Born-Oppenheimer correction, and $E_\mathrm{Rel.}$ denotes the scalar relativistic corrections to the energy based sum of the mass-velocity, one and two-electron Darwin terms. Details on the extrapolation schemes used can be found in Reference \citenum{harding_high-accuracy_2008}. The CCSDT(Q) calculation was performed using the MRCC program interfaced with CFOUR. \cite{kallay_mrcc_2017}

%%%%%%%%%%%%%%%%%%%%%%%%%%%%%%%%%%%%%%%%%%%%%%%%%%%%%%%%%%%%%%%%%%%%%%%%%%%%%%%%%%%%%%%%%
%%%%%%%%%%%%%%%%%%%%%%%%%%%  results                   %%%%%%%%%%%%%%%%%%%%%%%%%%%%%%%%%%
%%%%%%%%%%%%%%%%%%%%%%%%%%%%%%%%%%%%%%%%%%%%%%%%%%%%%%%%%%%%%%%%%%%%%%%%%%%%%%%%%%%%%%%%%

\section{Results}

\begin{figure*}[ht]
\centering
  \includegraphics[width=\textwidth]{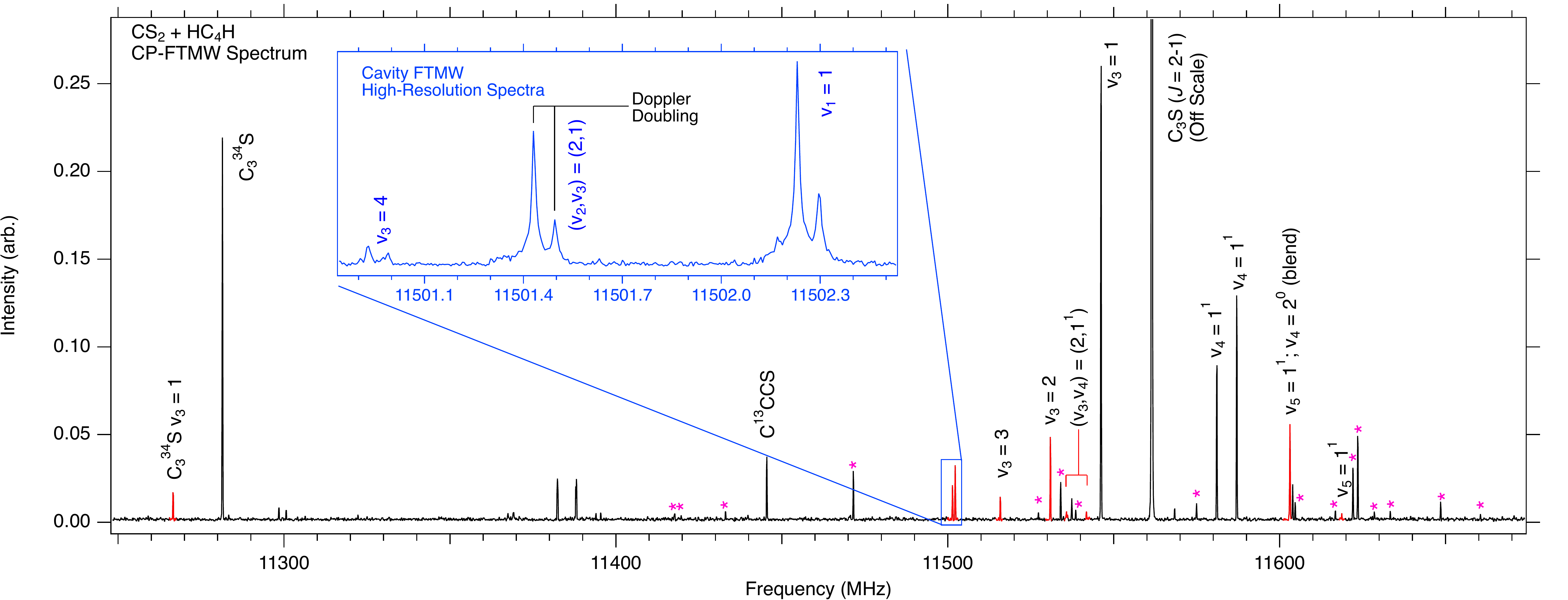}
  \caption{A portion of the  CP-FTMW spectrum obtained between 7.5 and 18\,GHz, through an electrical discharge of dilute \ce{CS2} + \ce{HC4H} in Ne/Ar, showing features near the $J=2 - 1$ transition of \ce{C3S}.  All \ce{C3S} lines that have been assigned from the present analysis are labeled in black, with  newly identified transitions highlighted in red.  The intensity of the ground state \ce{C3S} line has been truncated in this plot, to more easily illustrate less intense features, many of which are observed with high SNR; its peak value (4.1) is roughly 15 times that of the next strong feature in the frequency range. Lines that remain unidentified  are marked with asterisks. The inset shows a high-resolution cavity FTMW spectrum of several closely-spaced transitions; owing to the Doppler effect, each line in this spectrum consists of a closely-spaced doublet. 
  }
  \label{c3s_zoom}
\end{figure*}

Although there is considerable variance in the chemical richness of the four discharge mixtures (Table \ref{known_molecules}), all produce molecules of astronomical interest.  While \ce{CS2} alone only yields the shortest carbon-sulfur chains (\ce{C2S} to \ce{C4S})  in detectable abundances, evidence was found for more than a dozen acetylenic free radicals, carbenes, and methyl polyynes in the \ce{HC4H} discharge, nearly all of which have been detected in space. 
%Table \ref{new_molecules} delineates the newly-identified states of \ce{C2S} and \ce{C3S} while Table~\ref{known_molecules} provides a complete list of species detected in the different experimental spectra.  %
Addition of a hydrocarbon to a \ce{CS2} discharge results in a plethora of carbon and sulfur compounds, of which 36\% have already been observed in space. As illustrated in Figure~\ref{c3s_zoom} and Tables \ref{known_molecules}--\ref{known_molecules_on}, one of the most striking features in the \ce{CS2}/hydrocarbon discharges is the remarkably high SNR of small reactive species such as \ce{C3S}, and the large number of newly identified lines close in frequency (within a few per cent) to these strong features. Indeed, the SNR is high enough for \ce{C3S} that we were able to observe \ce{CCC^{33}S} in natural abundance in both the \ce{CS2 + HC4H} and \ce{CS2 + HCCH} spectra, and assign its rotational spectrum for the first time (see Tables~\ref{c3s_constants}~\&~\ref{ccc33s_freqs}).  Due to the high sensitivity of the measurements, evidence was routinely found for common contaminants such as \ce{SO2} and OCS in our chirped-pulse spectra. In particular, the \ce{CS2 + HCCH} mixture experiment was performed immediately after an experiment using \ce{SO2} as a precursor, yielding strong lines of this species, its vibrational satellites, and isotopologues in the broadband spectrum. It is also worth noting a weak \ce{C3S} line was detected in the CP-FTMW spectrum nominally containing only \ce{HC4H} and carrier gas, a testimony to the ease with which \ce{C3S} can be produced even when trace quantities of \ce{CS2} are apparently present.

\begin{center}
    
\begin{table}[p!]
\centering
  %\scriptsize
  {\fontsize{7}{8}\selectfont
  \caption{Pure hydrocarbon and sulfur-containing species identified in each reaction mixture whose frequencies were known prior to present work. Numbers  in the Table represent the signal-to-noise ratio of the strongest line assigned for each species.}
  \label{known_molecules}
  \begin{tabular}{l cccc} 
    \toprule
   Species$^a$ 		  				&	\ce{CS2 + HC4H}	&	 \ce{CS2 + HCCH}	&	 \ce{CS2}	&	\ce{HC4H}\\
    \midrule	
	\ce{C2S}     				& 161 	& 75	& 19 	& 3 	\\
	 \hspace{1em} CC$^{34}$S	&  19 	& 3 	& 3 	&  		\\
	 \hspace{1em} $^{13}$CCS  	&   2 	&   	&   	&   	\\
	\hspace{1em} C$^{13}$CS  	&   2 	&		&		&		\\
	
	\vspace{-0.75em}	\\
	\midrule[0.25pt]
	
	\ce{C3S}					        & 2040 	& 598 	& 18 	& 18 	\\
	\hspace{1em} $v_1=1\: ^d$  	        & 16	& 		&		&		\\
	\hspace{1em} $v_4=1^1\: ^d$ 	    & 138   & 67 	&3	    & 2		\\
	\hspace{1em} $v_5=1^1\: ^d$ 	    &   3	& 17	& 2		& 2		\\
	\hspace{1em} $v_3=1$   	    	    & 130	& 76	&3 		& 2		\\
	%\hspace{1em} \textcolor{blue}{$(v_1,v_3)=(1,1)\: ^d$} & X	    & X	    &X 		& X		\\ % Cavity only
	
	\vspace{-0.5em}	\\
		
	\hspace{1em} CCC$^{34}$S	& 109 	& 23	& 2		&  2 	\\
	\hspace{1em} $^{13}$CCCS  	& 19  	& 3		&		&		\\
	\hspace{1em} C$^{13}$CCS  	& 18  	& 5		&		&		\\ 
	
	\vspace{-0.75em}	\\
	\midrule[0.25pt]
		
	\ce{C4S}             		& 302	& 19 	& 5		& 5  	\\ 
	\ce{C5S}              		& 125	& 103	&		& 4		\\ 
	\hspace{1em} \ce{C5 ^{34}S} & 7		&   	&   	& 		\\ 

	\vspace{-0.5em}	\\
	
	\ce{C6S} 					& 12	& 3		& 		&		\\
	\ce{C7S}    				& 20	& 12	&   	&		\\
	\ce{C8S}    				& 3		& 		&   	&		\\	
	
	\vspace{-0.75em}	\\
	\midrule[0.25pt]
	
	\ce{HC3S}	          		&  50   & 152 	&		& 3		\\
	\ce{HC4S}   				& 215   & 110	& 2		& 3		\\
	\ce{HC5S}   				& 23	& 13	&		& 5		\\
	\ce{HC6S}   				& 17	& 17	&		&		\\
	\ce{HC7S}   				& 2		& 2		&		&		\\
	\ce{HC8S}   				& 5		& 2		&		&		\\ 
	
	\vspace{-0.75em} \\
	\midrule[0.25pt]
	
	\ce{H2C2S}  				& 2   	& 5 	&		&		\\
	\ce{H2C3S}  				& 21 	& 35 	&		&		\\
	\ce{H2C4S}            		& 17	& 23 	&		&  		\\
	\ce{H2C5S}            		& 6		& 12	&		& 		\\
	\ce{H2C6S}    	    		&		& 3		&		&		\\ 
	
	\vspace{-0.75em}	\\
	\midrule[0.25pt]
	
	\ce{HCSC2H} 		   		& 14	& 7		&		&		\\
	\ce{HCSC4H} 		    	&  9	& 11	&		&		\\ 
	
	\vspace{-0.75em}	\\
	\midrule[0.25pt]
	
    $c$-\ce{C3H}  	    		& 13	& 23 	&		& 54	\\
	\ce{C4H}    				& 20	&		&		& 171	\\
	\ce{C5H}    				& 26	& 9 	&		& 230	\\
	\ce{C6H}    				& 3  	&		&		& 49	\\
	\ce{C7H}    				&  4	&		&		& 77	\\
	\ce{C8H}					&  3	&		&		& 60	\\
	\ce{C9H}					&		&		&		& 32	\\
	\ce{C10H}					&		&		&		& 7		\\
	\ce{C11H}					&		&		&		& 4		\\	
	
	\vspace{-0.75em}	\\
	\midrule[0.25pt]
	
	$c$-\ce{C3H2} 				&  8    & 372 	&		& 16	\\
        \hspace{1em} $v_6=1$ 	&  3    & 5		&		& 7		\\
        \hspace{1em} $v_3=1$ 	&       & 5		&		&  		\\
        \hspace{1em} $v_2=1$ 	&       & 3		&		&  		\\

	\vspace{-0.5em}	\\
	
        \hspace{1em} C$^{13}$CHCH $^b$&  & 3 	&   	&   	\\ 
	
	\vspace{-0.75em}	\\
	\midrule[0.25pt]
	
	\ce{C4H2}   				& 6		&		&		& 16	\\
	$l$-\ce{C5H2}				& 3		&		&		& 35 	\\
	$c$-\ce{C5H2}				& 16	& 3		&		& 70	\\
	\ce{C5H2} $^c$				&       & 		&		& 25	\\
	\ce{C6H2}   				&		&  		&		& 3		\\
	\ce{C7H2}   				&		&  		&		& 6		\\ 
	
	\vspace{-0.75em}	\\
	\midrule[0.25pt]
		
	\ce{CH3C2H}	  	    		& 17	& 		&		& 9		\\
	\ce{CH3C4H}	          		& 2		& 3		& 		& 21	\\ 
	
	\vspace{-0.75em}	\\
	\midrule[0.25pt]
	
	SH $^e$	          	    		& 2 	& 48	&		&		\\ 
	\hspace{1em} $v=1$ $^d$   		&  		& 10   	&		&	   	\\
    \bottomrule
  \end{tabular}
  }
  
  \footnotesize
    \smallskip
  \begin{minipage}{0.45\textwidth}
  $^a$ Main isotopologue in its ground vibrational state, unless otherwise noted.\\
  $^b$ Refers to the isotopic variant in which $^{13}$C has been substituted at one of the two equivalent carbon atoms.\\
  $^c$ Refers to the bent-chain isomer, i.e.~isomer 3 in Ref.~\citenum{Gottlieb:1998}.\\
  $^d$ Centimeter-wave lines observed for the first time, using rotational constants determined from previous works \cite{Tang:1995jr,Dudek:2017ij,MartinDrumel:2012br}.  See Tables \ref{c3s_cm_freqs} and \ref{sh_cm_freq} for a complete listing of observed transitions. \\
  $^e$ One additional hyperfine line reported compared to Ref. \citenum{Meerts:1975kq} (see Table \ref{sh_cm_freq}).
  \end{minipage}
\end{table}

\end{center}

Nearly all of the species observed in the \ce{CS2 + HC2H} spectrum are also found in the  \ce{CS2 + HC4H} discharge, with the latter mixture generally resulting in much stronger lines for most carbon-rich molecules (Table \ref{known_molecules}). Because this mixture yields the richest array of compounds, it is the main focus of the spectral analysis presented here.  In this spectrum, 59 unique variants (including isotopic and vibrational states) from 42 chemical species were assigned; among these are five vibrational satellite transitions of \ce{C3S} and SH, which were observed in the centimeter domain for the first time using previously reported rotational constants (see Tables \ref{known_molecules}, \ref{c3s_cm_freqs}, and \ref{sh_cm_freq}). Once transitions of these species were assigned in the CP spectra, exhaustive binary DR tests were performed on the remaining transitions to identify other features that might arise from a common carrier or carriers.  Many DR linkages were found, but most connect only two lines separated by $\sim$5.8\,GHz, implying a rotational constant close to that of \ce{C3S} ($\sim$2.9\,GHz), since this molecule has two transitions within the range of the CP-FTMW spectrum, one at 11.5\,GHz ($J = 2-1$) and another at 17\,GHz ($J = 3-2$),  both of which lie close in frequency to the new DR linkages.  Taken together, these findings strongly suggest the presence of many new vibrationally excited states of \ce{C3S}. Surprisingly, some of the unidentified lines are comparably in intensity to lines from the $v_5=1^1$ (CCS bending), $v_4=1^1$ (CCC bending) and $v_3=1$ states of \ce{C3S}, which were previously observed by Tang and Saito,\cite{Tang:1995jr} and Crabtree \textit{et al.} \cite{Crabtree:2016fj}, respectively (see Fig~\ref{c3s_zoom}). 

Working under the operative assumption that these unidentified lines arise from still other vibrationally excited states of \ce{C3S}, and that additional transitions should obey a simple linear-molecule progression, follow-up cavity measurements were undertaken to detect higher-frequency ($>$18~GHz) transitions. Because the predicted frequencies of the higher-$J$ lines are simply related to one another by ratios of integers, little search was required. On the basis of theoretical vibration-rotation constants calculated previously~\cite{Seeger:1994bp} and those in \S\ref{qcc}, nearly 20 series have now been assigned to vibrationally excited normal \ce{C3S} or one it is more abundant isotopic species. Table~\ref{new_molecules} summarizes the newly-assigned transitions, while the measured centimeter-wave frequencies are given in Table~\ref{c3s_cm_freqs}.

\begin{table*}[ht!]
\centering
\caption{New states of \ce{C2S}, \ce{C3S} and \ce{C4S} assigned in this work. Numbers in brackets indicate the signal-to-noise ratio of the strongest assigned transition of the state in the \ce{CS2 + HC4H} reaction mixture chirped-pulse spectrum. Remaining states were observed only in the cavity instrument.}
\label{new_molecules}
\small 
\begin{tabular}{D{=}{~=~}{-1}@{}@{}r  D{=}{~=~}{-1}@{\hspace{1mm}} r c c c c c c}
\toprule
\multicolumn{1}{c}{\hspace{5mm}\ce{C2S}} &   &       \multicolumn{6}{c}{\ce{C3S}  }                                                             & \multicolumn{2}{c}{\ce{C4S}}    \\ 
\cmidrule(l{2pt}r{2pt}){3-8} 
\cmidrule(l{2pt}r{2pt}){9-10}
\multicolumn{1}{c}{\hspace{5mm}main}  &              	&     \multicolumn{1}{c}{\hspace{4mm}main}  &        & \ce{^{13}CCCS}  &   \ce{C^{13}CCS}  &    C$_3 ^{\,34}$S & C$_3 ^{\,33}$S               & main         & C$_4 ^{\,34}$S\\
\midrule
v_1 = 1     & $[19]$   & v_2=1  &    &   $v_3=1$  &   $v_3=1\; [3]$   &     $v_3=1\; [8]$   & GS $[9]$	& $v_6=1^1\; [8]$      & GS $[7]$\\
v_2 = 1^1   &          &   v_3=2    & $[24]$      		&            &             &     $v_4=1\; [5]$   & 		& $v_7=1^1\; [5]$ &\\
v_2 = 2^0   &          &       v_3=3 &  $[7]$     			& &&\\
v_3 = 1     &  $[38]$		&        v_3=4 &  &&&\\
v_3 = 2     &      &       v_4=2^{0,2} & $[28]$    		&&& \\
v_3 = 3     & 	&       (v_1,v_3)=(1,2) & $[2]$    	&&&\\
(v_1,v_2) = (1,1^1) &	&       (v_2,v_3)=(1,1) & $[10]$	&&&\\
(v_1,v_3) = (1,1)   & $[6]$	&       (v_3,v_4)=(1,1^1) & $[10]$ 	&&& \\
(v_2,v_3) = (1^1,1) &   	&       (v_2,v_4)=(1,1^1) & $[2]$  	&&& \\
(v_2,v_3) = (2^0,1)	&	&         			&&&& \\
 
\bottomrule
\end{tabular}

  \smallskip
  \begin{minipage}{0.9\textwidth}
  Note: Transitions of \ce{C3S} in $v_1=1$, $v_4=1$, $v_5=1, 2$, and $(v_1,v_3)=(1,1)$ have also been observed for the first time in the centimeter domain using previously reported spectroscopic constants for these states\cite{Tang:1995jr,Dudek:2017ij}, see Table \ref{c3s_cm_freqs}.
  \end{minipage}
\end{table*}

The presence of vibrationally excited lines of \ce{C3S} in our spectra suggests that \ce{C2S} and even \ce{C4S} may be excited similarly.  Only the $J_N = 2_1 - 1_0$ transition of \ce{C2S} near 11\,GHz, however, lies in the frequency range of the CP-FTMW spectrum, with the next strong transition lying closer to 22\,GHz;  consequently, no new lines of \ce{C2S} can be identified by DR using only the CP-FTMW coverage.  Nevertheless, in CP-FTMW spectra where \ce{C3S} lines are strong, vibrationally excited lines typically fall within a few \% (i.e.~a few 100\,MHz near 11\,GHz) of the ground state (see Fig.~\ref{c3s_zoom}). 
To establish if some of the unidentified lines near 11\,GHz arise from vibrationally excited \ce{C2S}, surveys covering roughly $\pm2$\% in frequency around the $J_N = 1_2 - 2_1$ transition of the ground state  near 22\,GHz were subsequently performed using cavity FTMW spectroscopy.  Several unidentified lines were observed in this frequency region, and soon afterwards linked to low-frequency lines in the CP-FTMW spectrum by DR.  Still higher-frequency transitions were then measured with the cavity spectrometer up to 40\,GHz, and an additional line was often detected by DR spectroscopy between 40 and 60\,GHz.  On the basis of the close agreement between the measured lines and predictions from the vibration-rotation constants in \S\ref{qcc}, these lines have been assigned to the $\nu_1$ and $\nu_2$ stretching modes, two quanta of the $\nu_3$ bending mode, or some combination of the three.  Table~\ref{c2s_cm_freqs} summarizes the centimeter-wave measurements of the new vibrational states.  

Although no DR linkages implicate new vibrationally excited states of \ce{C4S}, careful inspection of the four rotational transitions that lie in our CP-FTMW spectra revealed a weak cluster of features displaced to slightly higher frequency compared to the ground state line for each transition.   Subsequent assays, chemical tests, and DR measurements established that these lines behave as the ground state, and on the basis of the vibration-rotation constants given in Table~\ref{alphas}, were assigned to the two lowest-frequency bending modes ($v_6=1$ and $v_7=1$); these transition frequencies are summarized in Table~\ref{c4s_cm_freqs}. Under optimized experimental conditions, lines of $v_6=1$ are roughly 50 times weaker than the same lines of the ground state, while those of $v_7=1$ are closer to a 100 times weaker, implying T$_{\rm vib}\sim 130$\,K and $\sim 40$\,K respectively. Despite the comparably high SNR of lines of \ce{C5S} (Table~\ref{known_molecules}), and some prior experimental work in the infrared,\cite{Thorwirth:2017fg}, no vibrationally excited lines were found for this species.

On the basis of the newly measured lines of vibrationally excited \ce{C2S} and \ce{C3S} at low frequency, predictions were then made between 253 and 280\,GHz, a frequency region that coincides with a survey previously performed in a low pressure, long-path dc glow spectrometer (\S\ref{mmw}).  Because the expected uncertainty in line positions obtained by extrapolation is only a few MHz at these frequencies, assignments of higher-$J$ transitions of vibrationally excited \ce{C2S} ($J' = 20, 21$) and \ce{C3S} ($J' = 44-48$) were fairly straightforward (see Tables \ref{c2s_mm_freqs} and \ref{c3s_mm_freqs}).  Fits that combine both sets of measurements were then performed for each new state using the CALPGM (SPFIT/SPCAT) suite of programs.\cite{Pickett:1991cv}  The best-fit constants are given in Tables~\ref{c2s_constants} and \ref{c3s_constants} for \ce{C2S} and \ce{C3S}, respectively; those for \ce{C4S} are summarized in Table~\ref{c4s_constants}.  The data set for many vibrational states is limited to centimeter-wave measurements.  In these cases, the centrifugal distortion constant $D$ was fixed to the value of the normal isotopic species.  

In addition to vibrational satellite lines of \ce{C2S} and \ce{C3S}, transitions from vibrationally excited CS, C$^{34}$S, $^{13}$CS, and C$^{33}$S were also identified in the course of our re-analysis of the millimeter-wave survey. These measurements  are summarized in Table~\ref{cs_vibstates}.

%%%%%%%%%%%%%%%%%%%%%%%%%%%%%%%%%%%%%%%%%%%%%%%%%%%%%%%%%%%%%%%%%%%%%%%%%%%%%%%%%%%%%%%%%
%%%%%%%%%%%%%%%%%%%%%%%%%%%  Discussion                %%%%%%%%%%%%%%%%%%%%%%%%%%%%%%%%%%
%%%%%%%%%%%%%%%%%%%%%%%%%%%%%%%%%%%%%%%%%%%%%%%%%%%%%%%%%%%%%%%%%%%%%%%%%%%%%%%%%%%%%%%%%

\section{Discussion}

\subsection{MST as a stepping stone for millimeter-wave assignment and astrophysical implications}
The present work demonstrates a simple but highly useful aspect of MST -- the ability to rapidly and confidently identify new vibrational satellite transitions of abundant, well-studied molecules in a reaction mixture containing familiar and transient species.  As demonstrated here, an electrical discharge of two small-molecule precursors, \ce{CS2} and either \ce{HCCH} or \ce{HC4H}, produced a mixture of considerable complexity in which in excess of 70 unique chemical species, their more abundant isotopic variants and/or in their vibrationally excited states are present in detectable abundances; in total, 31 vibrational states or new isotopic were assigned for the first time.  Because this reaction screening technique can be implemented relatively easily, it may be an appealing alternative to traditional methods for detecting new species of plausible astronomical interest, their isotopic species, and in vibrationally excited states.  Although many transitions have been assigned, a large fraction ($\sim$40\%) remain unidentified, implying the discovery space for entirely new compounds is still sizable.  %for the record: 307 lines with SNR > 3 initially (after artifacts removed), 129 remain after all assignments

The use of CP-FTMW spectroscopy as a tool for molecular discovery was illustrated several years ago in the astronomical identification of ethaninimine (\ce{CH3CHNH})\cite{Loomis:2013fs} and E-cyanomethanimine (HNCHCN).\cite{Zaleski:2013bc} In that study,  a nearly identical discharge source was employed, and acetonitrile (\ce{CH3CN}) and ammonia (\ce{NH3}) were used as precursors. By directly comparing cm-wave CP-FTMW spectra to a frequency-coincident molecular line survey of the Sagittarius B2(N) star-forming region, several frequency coincidences were found, strongly suggesting a common carrier; subsequent laboratory work ultimately established the presence of both species in the interstellar medium (ISM) for the first time.  Both  \ce{CH3CHNH} and HNCHCN, however, were studied at least to some extent in previous microwave investigations. \cite{Lovas:1980jc,Brown:1980ga,Takano:1990hd}  In this sense, MST extends previous efforts by its ability to systematically and rapidly identify lines that arise from a unique species, regardless of whether the identity of the carrier is known from prior work.  In combination with theoretical calculations and other tests and assays, it is then often possible to deduce the elemental composition and structure of the carrier, as recent work on the isomers of \ce{H2C5O} and other long-chain cumulenones demonstrates.\cite{mccarthy:154301} 

By performing laboratory measurements at centimeter wavelengths with a jet source, where the detection sensitivity and the spectral resolution are both very high, and spectral confusion is not an issue,  measurements can be extended with little uncertainty to millimeter-wavelengths, where powerful millimeter-wave interferometers such as ALMA operate.  Because spectral confusion is also much more common in laboratory spectra at these wavelengths, there is great practical utility in using microwave data as a `stepping stone' to higher frequencies.  The identification and assignment of a significant number of new transitions of \ce{C2S} and \ce{C3S} in legacy spectra from our laboratory are but one such example.  Because the fits span  both high and low-frequency data, transitions at intermediate frequencies can be trivially predicted with high accuracy.

Finally, the fairly exhaustive analysis of the microwave spectra by MST enables a comprehensive characterization of the chemical and physical processes that are operative in an electrical discharge starting with either \ce{CS2} alone, or in combination with a hydrocarbon, either \ce{HCCH} or \ce{HC4H}.  Several examples highlighting this point are given in the sections that follow, including the formation pathways of \ce{C2S} and \ce{C3S}, their vibrational excitation, and the abundances of long-chain molecules in comparison to the ISM.

\subsection{Formation of \ce{C2S} and \ce{C3S}}

Following the detection of the simplest C$_n$S thiocumulenes, \ce{C2S} and \ce{C3S}, in the ISM,\cite{Saito:1987fa,Yamamoto:1987jd} in the late 1980s, the formation pathways of these molecules and longer members has been a topic of considerable debate.  While ion-neutral reactions involving S$^+$ were originally believed to be sufficient to reproduce the observed abundances,\cite{Smith:1988dc} detailed modeling based on these reactions revealed a sizable discrepancy (of several orders of magnitude) between predicted and observed abundances in IRC+10216.\cite{Millar:1990yu,Cernicharo:1987jh}  Subsequent theoretical investigations suggested that neutral-neutral and radical-neutral reactions likely play a significant role,\cite{Yamada:2002gm,Petrie:1996yd} a supposition which has recently been supported by observations of unequal $^{13}$C ratios in \ce{C2S} and \ce{C3S} towards the cold, dark molecular cloud TMC-1.\cite{Sakai:2007ud} Indeed, from the observed [C$^{13}$CS]/[$^{13}$CCS] abundance ratio of $\sim$4 in the cold molecular clouds TMC-1 and L1521E,  Sakai et al.\cite{Sakai:2007ud} concluded that reactions involving S/S$^+$ are not the main pathway to \ce{C2S}, and this species is instead likely formed from two reactants, each of which contributes a C atom.  An analogous argument has been put forth for \ce{C3S}~\cite{sakai:9831}.

In the laboratory, it is well established from prior experiments that, as in the interstellar medium, S and CS are the major fragmentation products in an \ce{CS2} electrical discharge\cite{seaver:63}, and that the fractional ionization is very low, of order $10^{-5}$ or less, when a heavy inert atom such Ne and Ar is used as the buffer gas~\cite{lattanzi:1717}. As a result, the steady-state abundance of S$^+$ should be several orders of magnitude lower than atomic S, and the ion-molecule reactions probably contribute little to the operative chemistry.

Reactions between neutral and radical species are instead expected to dominate, where hydrocarbon fragments are likely formed via several competing reactions:
    \begin{eqnarray}
        \ce{HCCH}   &\rightarrow& 2\ \ce{CH}		\\
                    &\rightarrow& \ce{C2H} + \ce{H}	\\
        \ce{HC4H}   &\rightarrow& 2\ \ce{C2H}		\\
                    &\rightarrow& \ce{C4H} + \ce{H}	\\
                    &\rightarrow& \ce{C3H} + \ce{CH}
    \end{eqnarray}

In our laboratory experiments, formation pathways for \ce{C2S} include:
\begin{eqnarray}
\ce{CH} + \ce{CS} &\rightarrow& \ce{C2S} + \ce{H} \label{eqn6}\\
\ce{C2H} + \ce{S} &\rightarrow& \ce{C2S} + \ce{H} \label{eqn7}
\end{eqnarray}
\noindent In fact, Eq.~\ref{eqn6} is thought to be the most probable route to produce \ce{C2S} in TMC-1.\cite{Sakai:2007ud}  Electrical discharge sources are notorious for their lack of specificity and rapid isotopic scrambling, and so we carried out \ce{C2S} isotopic experiments with acetylene and $^{13}$\ce{CS2} in place of \ce{CS2}, the results of which point to the importance of Eq.~\ref{eqn7}.  Under our experimental conditions, relatively little $^{13}$C enhancement is observed for either $^{13}$CCS or C$^{13}$CS, despite strong lines of normal CCS, an indication that the acetylenic unit remains largely intact during molecule formation, and that $^{13}$C from $^{13}$\ce{CS2} serves largely as a spectator.  This finding is consistent with earlier laboratory studies by Ikeda \textit{et al.}~\cite{Ikeda:1997ty} who first reported the rotational spectrum of $^{13}$CCS and C$^{13}$CS.  In that study, they concluded the C$\equiv$C bond in acetylene did not cleave efficiently, as the use of an enriched sample of H$^{13}$C$^{13}$CH did not result in stronger lines of either of the two $^{13}$C species. In fact, lines of the double-substituted species $^{13}$C$^{13}$CS were readily observed instead.

The formation of \ce{C3S} through the radical-radical recombination reaction
\begin{equation}
\ce{C2H} + \ce{CS} \rightarrow \ce{C3S} + \ce{H}\\
\label{eqn8}
\end{equation}
\noindent would appear to be a particularly promising route to form \ce{C3S} in our discharge experiments, since both radicals are known to be produced in high abundance from their respective precursors, and because radical-radical reactions are normally exothermic and barrierless. If a major pathway, it follows that the \ce{C-S} bond should remain intact during molecule formation. Analogous experiments to those performed on \ce{C2S} with $^{13}$\ce{CS2}, however, suggest a different pathway to \ce{C3S} than Eq.~\ref{eqn8}.  Under a wide range of conditions, including low concentrations of both precursors and very low discharge voltages, $^{13}$C insertion appeared to occur facilely but with little selectively, with C$^{13}$CCS or $^{13}$CCCS only at most a factor of two less abundant than CC$^{13}$CS. Equally surprising was the presence of strong lines of normal \ce{C3S} under the same conditions.  
Taken together, these findings suggest that: (1) \ce{C3} or a closely-related species such as \ce{C3H} radical serves as a key reaction partner, but one that must be formed via a cyclic intermediate or transition state so as to produce a nearly statistically distribution of $^{13}$C in the carbon chain; and (2)~a subsequent reaction with free sulfur then yields \ce{C3S}. Previous experimental and theoretical studies conclude that the reaction:
\begin{equation}
   \ce{C3}+\ce{H2} \rightarrow \ce{c-C3H2} 
\end{equation}
 is the most energetically stable product channel starting from HCCH and either C($^1$D) or C($^3$P).  This pathway is barrierless for C($^1$D) insertion, and while the same reaction with C($^3$P) is spin-forbidden, it is still thought to proceed efficiently via intersystem crossing~\cite{casavecchia:271}. If relevant to our discharge chemistry, this reaction may also help explain why $^{13}$C isotopic scrambling in longer hydrocarbon chains such as \ce{C5H}, \ce{C6H}, and \ce{C7H} is so prevalent~\cite{mccarthy:174308}.

\subsection{Vibrational Excitation}

Molecules produced in high abundance in our discharge source frequently possess some degree of vibrational excitation, and \ce{C2S} and \ce{C3S} are no exception. As demonstrated in earlier studies~\cite{Sanz:2003iya,Sanz:2005kc}, rotational satellite transitions from vibrationally excited states are commonly observed for small abundant molecules despite the very low rotational temperature in the jet expansion (typically $\sim$1-3\,K). The vibrational distribution is highly non-thermal due to the competition between excitation, which includes collisions with electrons having an average kinetic energy of 1--3\,eV and the excess internal energy that the molecule may possess as a result of formation, and relaxation which is dominated by collisional cooling.  As a consequence of these two competing factors, vibrational modes that have frequencies much below room temperature can be efficiently cooled on the timescale of the expansion, while one or more modes usually lying slightly above room temperature are "frozen out"; lines from high-frequency stretches are uniformly weak, ostensibly because the density of states increases quickly with vibrational energy, and there is a commensurate increase in the rate of internal vibrational relaxation (IVR).  Given the complexity of IVR processes in polyatomic molecules, it is almost impossible to predict details of this behavior in advance, especially when the formation pathway and internal energy distribution of the molecule are rarely known.  Nevertheless, the degree of vibrational excitation in our discharge experiments tends to fall off quickly with increasing size of the molecule, and is only infrequently observed for carbon chains with more than about six heavy (carbon-like) atoms.  For \ce{C5S}, for example, no lines that could be attributed to vibrationally excited states were identified in our CP-FTMW spectra, despite detection of lines of C$_5 ^{34}$S in natural abundance.

\begin{figure}[ht]
%\centering
  \includegraphics[width=0.5\textwidth]{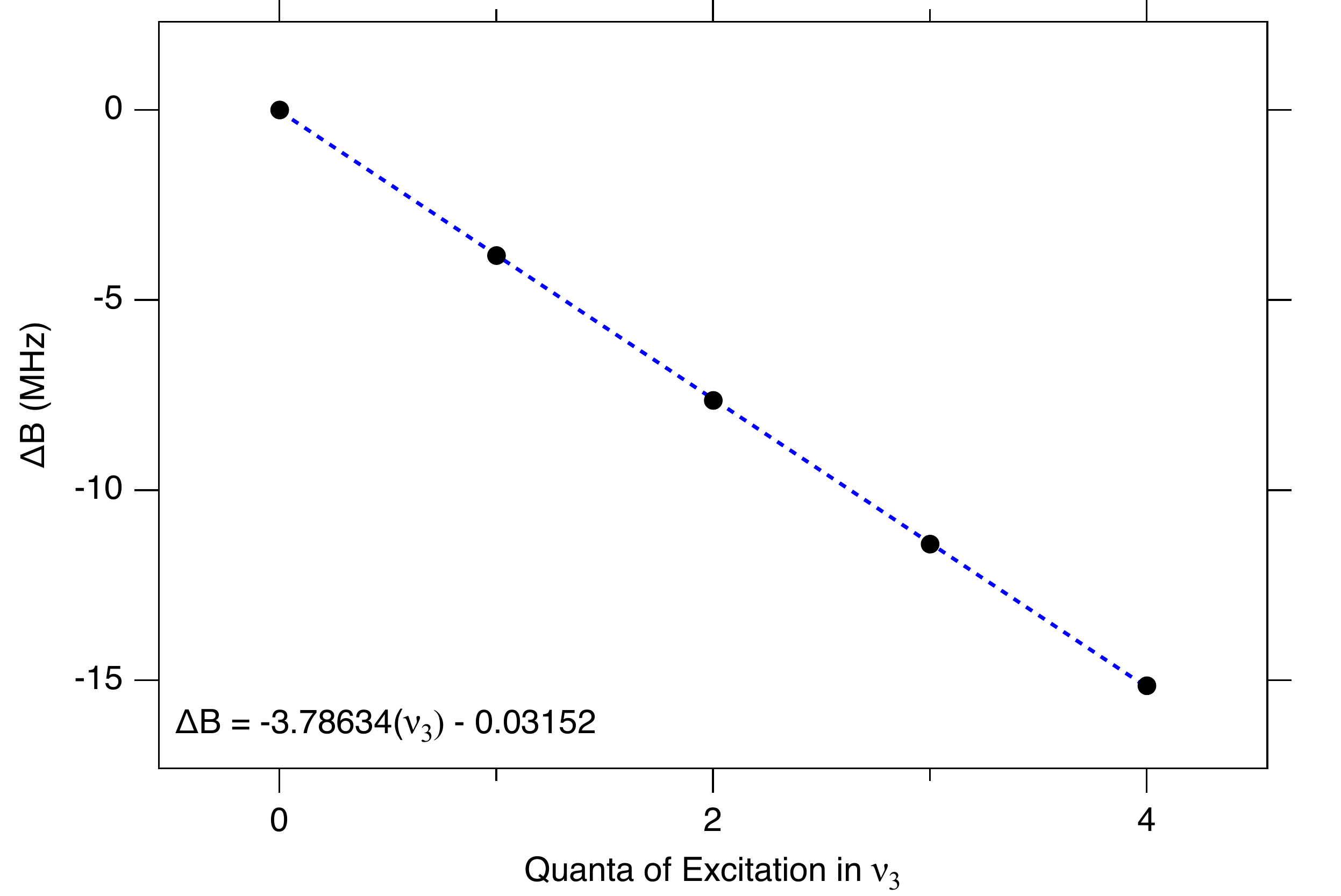}
  \caption{Differences between the rotational constants of \ce{C3S} ($\Delta B= B_v -B_0$) in the ground and vibrationally excited levels of the $\nu_3$ mode versus vibrational quantum number, $v_3$. The experimental value for $\alpha_3$ was determined from a least-squares
optimization (blue trace, $r$= 0.99997; see text).}
\label{alpha_plot}
\end{figure}

Because \ce{C3S} has a $^1\Sigma^+$ ground state with a harmonically-related transitions, analysis of its rotational satellite transitions is fairly straightforward, and consequently many vibrational states from either the normal or its rare isotopic species were assigned for the first time (Table \ref{new_molecules}).   As indicated in Fig.~\ref{c3s_zoom}, transitions from $\nu_3$, the lowest-energy stretching mode, and the $\nu_4$ bend are particularly intense, regardless of whether acetylene or diacetylene is used as the hydrocarbon source.  Excitation of $\nu_3$ is especially prominent in that states with as much as four quanta ($E\approx$2892~cm$^{-1}$) have been assigned. 
Table~\ref{alphas} provides a comparison of the experimentally-derived vibration-rotation constants $\alpha_i$ to those predicted for \ce{C3S} and \ce{C2S}, in which  $\alpha_i$ were obtained by differences of the rotational constants with respect to the ground state using the expression:
\begin{equation}
\Delta B= B_0-B_{v_i} \approx v_i \cdot \alpha_i		
\label{eqn9}
\end{equation}
\noindent where $i$ refers to the mode $\nu_i$.  For the $\nu_3$ mode, in which transitions from multiple quanta were observed, $\alpha_3$ was derived from linear regression as a function of the vibrational quantum number, $v_3$ (Fig.~\ref{alpha_plot}) this analysis yields  a precise, best-fit value within 10\% of that predicted from Seeger \textit{et al.}\cite{Seeger:1994bp}.  

\begin{figure}[t!]
\centering
  \includegraphics[width=0.5\textwidth]{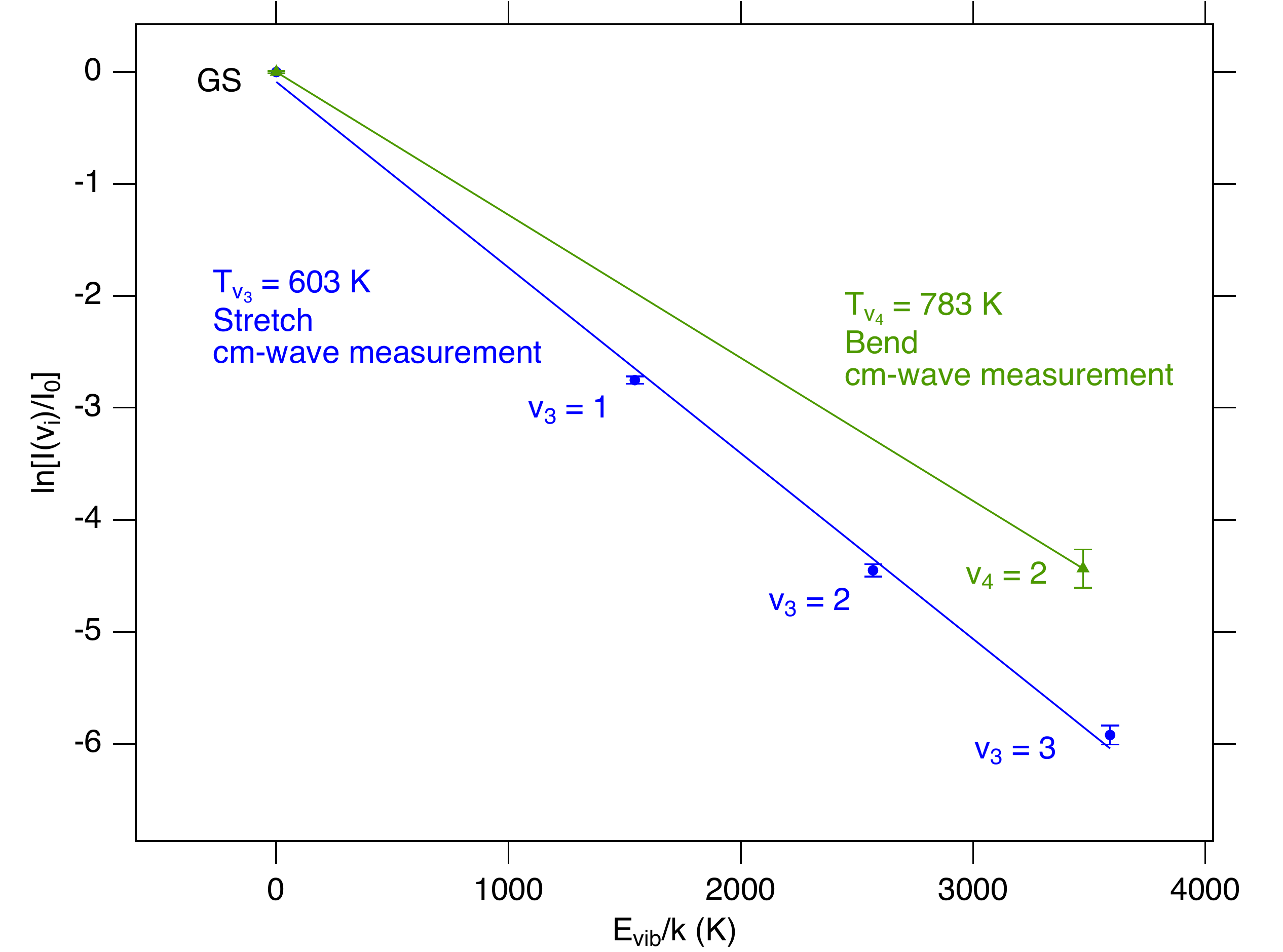}
  \caption{Vibrational temperature diagram of $\nu_4$ and $\nu_3$ of \ce{C3S} in the cm-wave CP-FTMW spectra.  Error bars (1$\sigma$) arise from uncertainties in the intensities that were derived from Gaussian fits to the line profiles.}
  \label{vib_temps}
\end{figure}

How excess energy is partitioned among the five vibrational modes of \ce{C3S} may provide clues as to its formation mechanism.  It is perhaps not surprising that significant energy would be concentrated in the $\nu_3$ mode (Fig.~\ref{vib_temps}) because it involves motion of the C--S unit, and the reaction of S atom with \ce{C3} (or a hydrocarbon fragment with the same number of C atoms) has been implicated as an important pathway to form \ce{C3S}: based on our calculations with HEAT345(Q) thermochemistry, the association of \ce{C3 + S -> C3S} is highly exothermic (-594.8\,kJ~mol$^{-1}$). Regarding how the reaction may occur, we can attempt to speculate on a mechanism based on the observed partitioning of this excess energy. As seen in Figure \ref{vib_temps}, vibrational temperatures of order 700\,K are found for $\nu_3$ and $\nu_4$.  This effective temperature is remarkably similar to those previously derived in our laboratory for chains such as \ce{HC3N}. \cite{Sanz:2005kc} Assuming the reaction proceeds barrierlessly, as is typical for radical-radical recombination reactions, the excess energy should be partitioned statistically. While the vibrational temperatures of $\nu_3$ and $\nu_4$ are comparable and therefore suggestive of a statistical distribution of states, the remaining vibrational modes ($\nu_1$, $\nu_2$, and $\nu_3$) are also observed, but are much less intense. However, as mentioned earlier in this section, vibrations with frequencies that deviate from room temperature significantly are generally cooled efficiently, or scrambled through IVR. Given the large body of experimental measurements that are now available for \ce{C3S}, it may be feasible to construct an accurate global potential energy surface, and trajectory simulations may prove enlightening.

\begin{table}[ht!]
\small \centering
\caption{\label{alphas}Theoretical and experimental vibration-rotation ($\alpha_i$) interaction constants of \ce{C2S}, \ce{C3S}, and \ce{C4S} (in MHz).}
\begin{tabular}{cc D{.}{.}{-1} D{.}{.}{-1}}
\toprule
Iso.         & Vib.                      &	\multicolumn{2}{c}{$\alpha_i$} 	\\
\cmidrule{3-4}
Species      &  Mode                      &	\multicolumn{1}{c}{Theory $^a$ }		&	 \multicolumn{1}{c}{Experiment} \\
\midrule
\ce{C2S}    &    $v_1$ 		              &    46.6         &   47.1             \\
            &    $v_2$ 		              &    19.0         &   20.0           \\
            &    $v_3$		              &   -27.8         &  -29.7              \\
\vspace{-0.5em} \\
\ce{C3S}    &    $v_1$ 		              &   15.2\; ^b     &  14.8\: ^c          \\
            &    $v_2$ 		              &   11.1\: ^b     &  11.2             \\
            &    $v_3$		              &    4.2\: ^b     &     3.79           \\
            &    $v_4$		              &   -5.2\: ^b     &     -5.6\: ^d         \\
            &    $v_5$		              &  -11.2\: ^b     &    -12.4\: ^d          \\
\vspace{-0.5em} \\
\ce{C4S}    &    $v_1$ 		              &   6.81         &         \\
            &    $v_2$ 		              &   4.86         &             \\
            &    $v_3$		              &   3.52         &           \\
            &    $v_4$		              &   1.34         &          \\
            &    $v_5$		              &  -1.88         &            \\
            &    $v_6$		              &  -2.62         &   -2.61            \\
            &    $v_7$		              &  -3.28         &   -3.07           \\
            \bottomrule
\end{tabular}

    \smallskip
    \begin{minipage}{0.4\textwidth}
        $^a$ Unless otherwise noted, the theoretical constants were calculated at the fc-CCSD(T)/cc-pVDZ level of theory.\\
        $^b$ Ref.~\citenum{Seeger:1994bp}\\
        $^c$ Ref.~\citenum{Dudek:2017ij}.\\
        $^d$ Ref.~\citenum{Tang:1995jr}.
    \end{minipage}
\end{table}

\ce{C2S} possesses a $^3\Sigma^-$ ground state with a very large spin-spin constant ($\lambda$),  which makes a detailed analysis of its vibrational excitation more challenging because the transitions are not strictly related to one another by ratios of integers at low $J$. Nevertheless, rotational lines from 10 vibrational states have been assigned either in our CP-FTMW spectra or in subsequent cavity searches at higher frequency, guided by theoretical calculations of the vibration-rotation coupling constants (\S\ref{qcc}).  These include either the fundamental or overtone of each mode or some combination of the three.  This degree of excitation appears fairly common for small molecules, e.g.~triatomics, that are either produced or subjected to an electrical discharge.  Strong satellites transitions are frequently observed in most or all of the vibrational modes, presumably because coupling between modes is relatively inefficient due to a low density of states, and because rotational spectra of many small molecules can frequently be observed with very high SNRs. Although  evidence was also found for all five vibrational modes of \ce{C3S}, several are very weak in our spectra.  In contrast, only two modes of \ce{C4S} and no modes of \ce{C5S} were found under the same experimental conditions, strongly suggesting that IVR plays a prominent role in rapidly and efficiently dissipating internal energy.  The experimental $\alpha$'s derived for each state of \ce{C2S} compare quite favorably to those calculated (Table~\ref{alphas}), indicating that the current theoretical treatment is adequate.

\subsection{Production of Longer Sulfur-Terminated Carbon-Chains}

The longest carbon chains detected in the ISM are \ce{C8H}/\ce{C8H^-}, \ce{HC9N}, \ce{C3S}, and \ce{HC5O}, for hydrogen, nitrogen, sulfur, and oxygen-bearing species, respectively.\cite{Cernicharo:1996kd,Remijan:2007vp,Brunken:2007yd,Broten:1978iu,Loomis:2016jsa,Matthews:1984hx,McGuire:2017ud}  In TMC-1, for example, lines of single $^{13}$C-\ce{C3S} have been reported, but no evidence has been found for \ce{C5S}, despite the availability of precise laboratory rest frequencies~\cite{Kasai:1993ds} for more than 20 years, and construction of new single-dish telescopes, such as the 100\,m GBT telescope, which have even greater collecting area.  A tentative detection has been reported at higher frequencies in IRC+10216.\cite{Agundez:2014gm}  In contrast, radio lines of neutral and negatively-charged acetylenic chains as long as \ce{C8H}/\ce{C8H^-} have been found in TMC-1.  Under our experimental conditions, spectra of chains as long as \ce{C7S} and \ce{C8H} are simultaneously observed in CP-FTMW spectra, suggesting there is no obvious kinetic or thermodynamic obstacle to formation of sulfur-terminated chains beyond \ce{C3S}. Rather, the stability of \ce{C3S} combined with the well-known depletion of sulfur in dense, cold molecular clouds (at the level of 99.9\% relative to the cosmic value\cite{tieftrunk:579}) point to elemental abundance as a mitigating factor in the production of longer sulfur-terminated chains in this source.

\subsection{Further Analysis}

Despite attempts to comprehensively analyze the spectra of hydrocarbon-sulfur discharges, about 40\% of the lines in our CP-FTMW spectra with a SNR in excess of 3 remain unassigned; the strongest of these are observed with a SNR close to 100.  Undoubtedly some fraction of these lines arise from still higher quanta or combination modes of normal and isotopic \ce{C2S}, \ce{C3S} and \ce{C4S} or other abundant  discharge molecules, such as \ce{HC3S}, HC(S)C$_2$H, etc.~while others may arise from relatively light molecules such as $c$-\ce{C3H2}, which only possess a single transition in the 8--18\,GHz frequency range of the CP-FTMW spectrometer.  Because these features will have no DR matches in the measurement range, further analysis and assignment is challenging.  Some of the strongest unidentified lines in the \ce{HC4H} discharge, for example, were recently assigned to one or more quanta in the $v_6$ mode of $c$-\ce{C3H2} in an unrelated study~\cite{ISMS}. To more easily identify light molecules and more routinely detect multiple DR linkages of the same molecule,  a  three-band system for CP-FTMW operating between 2 and 26.5\,GHz will soon be implemented.

\section{Conclusions}

An extensive MST analysis of several hydrocarbon/\ce{CS2} discharges has revealed the presence of many vibrationally excited states of \ce{C2S}, \ce{C3S,} and \ce{C4S}.  Subsequent analysis using new or existing theoretical vibration-rotation constants has enabled a total 27 new vibrational states of the three chains to be assigned; in combination with previously identified species, 90 unique products were assigned in the \ce{CS2 + HC4H} discharge. Predictions from the centimeter-wave data allowed previously unidentified lines of these species in archival millimeter-wave data to be identified and assigned with confidence.  In this way, complete and accurate spectral catalogs of species over the entire range of interest to radio astronomers can be compiled for subsequent use in analyzing complex interstellar mixtures.  This approach is particularly appealing for vibrationally excited species, which serve as excellent probes of physical conditions in the ISM, and offer access to different spatial scales than their ground vibrational state counterparts, particularly in regions where the lowest-energy species is optically thick.  Finally, isotopic spectroscopy using $^{13}$CS$_2$ indicated that the dominant formation pathway for \ce{C2S} in our laboratory discharge likely proceeds through \ce{C2H} + S, while \ce{C3S} appears to be formed from a cyclic intermediate, since $^{13}$C is found to be nearly randomly distributed along the chain.

\section*{Acknowledgements}

The work was supported by NSF grant AST-1615847. The National Radio Astronomy Observatory is a facility of the National Science Foundation operated under cooperative agreement by Associated Universities, Inc.  Support for B.A.M. was provided by NASA through Hubble Fellowship grant \#HST-HF2-51396 awarded by the Space Telescope Science Institute, which is operated by the Association of Universities for Research in Astronomy, Inc., for NASA, under contract NAS5-26555. 

\balance

%%%%%%%%%%%%%%%%%%%%%%%%%%%%%%%%%%%%%%%%%%%%%%%%%%%%%%%%%%%%%%%%%%%%%%%%%%%%%%%%%%%%%%%%%
%%%%%%%%%%%%%%%%%%%%%%%%%%%  end of main text          %%%%%%%%%%%%%%%%%%%%%%%%%%%%%%%%%%
%%%%%%%%%%%%%%%%%%%%%%%%%%%%%%%%%%%%%%%%%%%%%%%%%%%%%%%%%%%%%%%%%%%%%%%%%%%%%%%%%%%%%%%%%

%%%REFERENCES%%%

\bibliography{bibliography,random,c2s_add_refs} %You need to replace "rsc" on this line with the name of your .bib file

\providecommand*{\mcitethebibliography}{\thebibliography}
\csname @ifundefined\endcsname{endmcitethebibliography}
{\let\endmcitethebibliography\endthebibliography}{}
\begin{mcitethebibliography}{79}
\providecommand*{\natexlab}[1]{#1}
\providecommand*{\mciteSetBstSublistMode}[1]{}
\providecommand*{\mciteSetBstMaxWidthForm}[2]{}
\providecommand*{\mciteBstWouldAddEndPuncttrue}
  {\def\EndOfBibitem{\unskip.}}
\providecommand*{\mciteBstWouldAddEndPunctfalse}
  {\let\EndOfBibitem\relax}
\providecommand*{\mciteSetBstMidEndSepPunct}[3]{}
\providecommand*{\mciteSetBstSublistLabelBeginEnd}[3]{}
\providecommand*{\EndOfBibitem}{}
\mciteSetBstSublistMode{f}
\mciteSetBstMaxWidthForm{subitem}
{(\emph{\alph{mcitesubitemcount}})}
\mciteSetBstSublistLabelBeginEnd{\mcitemaxwidthsubitemform\space}
{\relax}{\relax}

\bibitem[Taatjes \emph{et~al.}(2008)Taatjes, Hansen, Osborn,
  Kohse-H{\"o}inghaus, Cool, and Westmoreland]{Taatjes:2008cl}
C.~A. Taatjes, N.~Hansen, D.~L. Osborn, K.~Kohse-H{\"o}inghaus, T.~A. Cool and
  P.~R. Westmoreland, \emph{Phys. Chem. Chem. Phys.}, 2008, \textbf{10},
  20--34\relax
\mciteBstWouldAddEndPuncttrue
\mciteSetBstMidEndSepPunct{\mcitedefaultmidpunct}
{\mcitedefaultendpunct}{\mcitedefaultseppunct}\relax
\EndOfBibitem
\bibitem[Semmelroch and Grosch(1995)]{Semmelroch:1995hy}
P.~Semmelroch and W.~Grosch, \emph{LWT - Food Science and Technology}, 1995,
  \textbf{28}, 310--313\relax
\mciteBstWouldAddEndPuncttrue
\mciteSetBstMidEndSepPunct{\mcitedefaultmidpunct}
{\mcitedefaultendpunct}{\mcitedefaultseppunct}\relax
\EndOfBibitem
\bibitem[Loomis \emph{et~al.}(2013)Loomis, Zaleski, Steber, Neill, Muckle,
  Harris, Hollis, Jewell, Lattanzi, Lovas, Martinez, McCarthy, Remijan, Pate,
  and Corby]{Loomis:2013fs}
R.~A. Loomis, D.~P. Zaleski, A.~L. Steber, J.~L. Neill, M.~T. Muckle, B.~J.
  Harris, J.~M. Hollis, P.~R. Jewell, V.~Lattanzi, F.~J. Lovas, O.~J. Martinez,
  M.~C. McCarthy, A.~J. Remijan, B.~H. Pate and J.~F. Corby, \emph{The
  Astrophysical Journal Letters}, 2013, \textbf{765}, L9\relax
\mciteBstWouldAddEndPuncttrue
\mciteSetBstMidEndSepPunct{\mcitedefaultmidpunct}
{\mcitedefaultendpunct}{\mcitedefaultseppunct}\relax
\EndOfBibitem
\bibitem[Kowalsick \emph{et~al.}(2014)Kowalsick, Kfoury, Robbat, Ahmed, Orians,
  Griffin, Cash, and Stepp]{kowaklick:230}
A.~Kowalsick, N.~Kfoury, A.~Robbat, S.~Ahmed, C.~Orians, T.~Griffin, S.~B. Cash
  and J.~R. Stepp, \emph{Journal of Chromatography A}, 2014, \textbf{1370}, 230
  -- 239\relax
\mciteBstWouldAddEndPuncttrue
\mciteSetBstMidEndSepPunct{\mcitedefaultmidpunct}
{\mcitedefaultendpunct}{\mcitedefaultseppunct}\relax
\EndOfBibitem
\bibitem[Teixeira and Rodrigues(2014)]{teixeira:9875}
M.~A. Teixeira and A.~E. Rodrigues, \emph{Industrial \& Engineering Chemistry
  Research}, 2014, \textbf{53}, 9875--9882\relax
\mciteBstWouldAddEndPuncttrue
\mciteSetBstMidEndSepPunct{\mcitedefaultmidpunct}
{\mcitedefaultendpunct}{\mcitedefaultseppunct}\relax
\EndOfBibitem
\bibitem[Daniel \emph{et~al.}(2011)Daniel, Tian, Xu, Wyszynski, Wu, and
  Huang]{Daniel:2011gl}
R.~Daniel, G.~Tian, H.~Xu, M.~L. Wyszynski, X.~Wu and Z.~Huang, \emph{Fuel},
  2011, \textbf{90}, 449--458\relax
\mciteBstWouldAddEndPuncttrue
\mciteSetBstMidEndSepPunct{\mcitedefaultmidpunct}
{\mcitedefaultendpunct}{\mcitedefaultseppunct}\relax
\EndOfBibitem
\bibitem[Rosatella \emph{et~al.}(2011)Rosatella, Simeonov, Frade, and
  Afonso]{Rosatella:2011ju}
A.~A. Rosatella, S.~P. Simeonov, R.~F.~M. Frade and C.~A.~M. Afonso,
  \emph{Green Chem.}, 2011, \textbf{13}, 754--40\relax
\mciteBstWouldAddEndPuncttrue
\mciteSetBstMidEndSepPunct{\mcitedefaultmidpunct}
{\mcitedefaultendpunct}{\mcitedefaultseppunct}\relax
\EndOfBibitem
\bibitem[Wu \emph{et~al.}(2009)Wu, Huang, Yuan, Zhang, and Wei]{Wu:2009ji}
X.~Wu, Z.~Huang, T.~Yuan, K.~Zhang and L.~Wei, \emph{Combustion and Flame},
  2009, \textbf{156}, 1365--1376\relax
\mciteBstWouldAddEndPuncttrue
\mciteSetBstMidEndSepPunct{\mcitedefaultmidpunct}
{\mcitedefaultendpunct}{\mcitedefaultseppunct}\relax
\EndOfBibitem
\bibitem[Togb{\'e} \emph{et~al.}(2014)Togb{\'e}, Tran, Liu, Felsmann,
  O{\ss}wald, Glaude, Sirjean, Fournet, Battin-Leclerc, and
  Kohse-H{\"o}inghaus]{TogbA:2014hm}
C.~Togb{\'e}, L.-S. Tran, D.~Liu, D.~Felsmann, P.~O{\ss}wald, P.-A. Glaude,
  B.~Sirjean, R.~Fournet, F.~Battin-Leclerc and K.~Kohse-H{\"o}inghaus,
  \emph{Combustion and Flame}, 2014, \textbf{161}, 780--797\relax
\mciteBstWouldAddEndPuncttrue
\mciteSetBstMidEndSepPunct{\mcitedefaultmidpunct}
{\mcitedefaultendpunct}{\mcitedefaultseppunct}\relax
\EndOfBibitem
\bibitem[Brown \emph{et~al.}(2008)Brown, Dian, Douglass, Geyer, Shipman, and
  Pate]{Brown:2008gk}
G.~G. Brown, B.~C. Dian, K.~O. Douglass, S.~M. Geyer, S.~T. Shipman and B.~H.
  Pate, \emph{Rev Sci Instrum}, 2008, \textbf{79}, 053103--14\relax
\mciteBstWouldAddEndPuncttrue
\mciteSetBstMidEndSepPunct{\mcitedefaultmidpunct}
{\mcitedefaultendpunct}{\mcitedefaultseppunct}\relax
\EndOfBibitem
\bibitem[Crabtree \emph{et~al.}(2016)Crabtree, Martin-Drumel, Brown, Gaster,
  Hall, and McCarthy]{Crabtree:2016fj}
K.~N. Crabtree, M.-A. Martin-Drumel, G.~G. Brown, S.~A. Gaster, T.~M. Hall and
  M.~C. McCarthy, \emph{J Chem Phys}, 2016, \textbf{144}, 124201--13\relax
\mciteBstWouldAddEndPuncttrue
\mciteSetBstMidEndSepPunct{\mcitedefaultmidpunct}
{\mcitedefaultendpunct}{\mcitedefaultseppunct}\relax
\EndOfBibitem
\bibitem[Garrod \emph{et~al.}(2008)Garrod, Weaver, and Herbst]{Garrod:2008tk}
R.~T. Garrod, S.~L.~W. Weaver and E.~Herbst, \emph{Astrophys J}, 2008,
  \textbf{682}, 283--302\relax
\mciteBstWouldAddEndPuncttrue
\mciteSetBstMidEndSepPunct{\mcitedefaultmidpunct}
{\mcitedefaultendpunct}{\mcitedefaultseppunct}\relax
\EndOfBibitem
\bibitem[Belloche \emph{et~al.}(2016)Belloche, M{\"u}ller, Garrod, and
  Menten]{Belloche:2016fm}
A.~Belloche, H.~S.~P. M{\"u}ller, R.~T. Garrod and K.~M. Menten,
  \emph{Astronomy {\&} Astrophysics}, 2016, \textbf{587}, A91--66\relax
\mciteBstWouldAddEndPuncttrue
\mciteSetBstMidEndSepPunct{\mcitedefaultmidpunct}
{\mcitedefaultendpunct}{\mcitedefaultseppunct}\relax
\EndOfBibitem
\bibitem[J{\o}rgensen \emph{et~al.}(2016)J{\o}rgensen, van~der Wiel, Coutens,
  Lykke, Muller, van Dishoeck, Calcutt, Bjerkeli, Bourke, Drozdovskaya,
  Fayolle, Favre, Garrod, Jacobsen, {\"O}berg, Persson, and
  Wampfler]{Jorgensen:2016cq}
J.~K. J{\o}rgensen, M.~van~der Wiel, A.~Coutens, J.~Lykke, H.~Muller, E.~van
  Dishoeck, H.~Calcutt, P.~Bjerkeli, T.~Bourke, M.~Drozdovskaya, E.~Fayolle,
  C.~Favre, R.~Garrod, S.~Jacobsen, K.~{\"O}berg, M.~Persson and S.~Wampfler,
  \emph{Astronomy {\&} Astrophysics}, 2016, \textbf{595}, A117\relax
\mciteBstWouldAddEndPuncttrue
\mciteSetBstMidEndSepPunct{\mcitedefaultmidpunct}
{\mcitedefaultendpunct}{\mcitedefaultseppunct}\relax
\EndOfBibitem
\bibitem[Cernicharo \emph{et~al.}(2013)Cernicharo, Daniel, Castro-Carrizo,
  Ag{\'u}ndez, Marcelino, Joblin, Goicoechea, and
  Gu{\'e}lin]{Cernicharo:2013cc}
J.~Cernicharo, F.~Daniel, A.~Castro-Carrizo, M.~Ag{\'u}ndez, N.~Marcelino,
  C.~Joblin, J.~R. Goicoechea and M.~Gu{\'e}lin, \emph{ApJ}, 2013,
  \textbf{778}, L25--6\relax
\mciteBstWouldAddEndPuncttrue
\mciteSetBstMidEndSepPunct{\mcitedefaultmidpunct}
{\mcitedefaultendpunct}{\mcitedefaultseppunct}\relax
\EndOfBibitem
\bibitem[Fortman \emph{et~al.}(2012)Fortman, McMillan, Neese, Randall, Remijan,
  Wilson, and De~Lucia]{Fortman:2012is}
S.~M. Fortman, J.~P. McMillan, C.~F. Neese, S.~K. Randall, A.~J. Remijan, T.~L.
  Wilson and F.~C. De~Lucia, \emph{J Mol Spectrosc}, 2012, \textbf{280},
  11--20\relax
\mciteBstWouldAddEndPuncttrue
\mciteSetBstMidEndSepPunct{\mcitedefaultmidpunct}
{\mcitedefaultendpunct}{\mcitedefaultseppunct}\relax
\EndOfBibitem
\bibitem[Cernicharo \emph{et~al.}(2010)Cernicharo, Waters, Decin, Encrenaz,
  Tielens, Ag{\'u}ndez, De~Beck, M{\"u}ller, Goicoechea, Barlow, Benz, Crimier,
  Daniel, di~Giorgio, Fich, Gaier, Garc{\'\i}a-Lario, de~Koter, Khouri, Liseau,
  Lombaert, Erickson, Pardo, Pearson, Shipman, S{\'a}nchez~Contreras, and
  Teyssier]{Cernicharo:2010es}
J.~Cernicharo, L.~B. F.~M. Waters, L.~Decin, P.~Encrenaz, A.~G. G.~M. Tielens,
  M.~Ag{\'u}ndez, E.~De~Beck, H.~S.~P. M{\"u}ller, J.~R. Goicoechea, M.~J.
  Barlow, A.~Benz, N.~Crimier, F.~Daniel, A.~M. di~Giorgio, M.~Fich, T.~Gaier,
  P.~Garc{\'\i}a-Lario, A.~de~Koter, T.~Khouri, R.~Liseau, R.~Lombaert,
  N.~Erickson, J.~R. Pardo, J.~C. Pearson, R.~Shipman, C.~S{\'a}nchez~Contreras
  and D.~Teyssier, \emph{A{\&}A}, 2010, \textbf{521}, L8\relax
\mciteBstWouldAddEndPuncttrue
\mciteSetBstMidEndSepPunct{\mcitedefaultmidpunct}
{\mcitedefaultendpunct}{\mcitedefaultseppunct}\relax
\EndOfBibitem
\bibitem[Cernicharo \emph{et~al.}(2008)Cernicharo, Gu{\'e}lin, Ag{\'u}ndez,
  McCarthy, and Thaddeus]{Cernicharo:2008wi}
J.~Cernicharo, M.~Gu{\'e}lin, M.~Ag{\'u}ndez, M.~C. McCarthy and P.~Thaddeus,
  \emph{The Astrophysical Journal}, 2008, \textbf{688}, L83\relax
\mciteBstWouldAddEndPuncttrue
\mciteSetBstMidEndSepPunct{\mcitedefaultmidpunct}
{\mcitedefaultendpunct}{\mcitedefaultseppunct}\relax
\EndOfBibitem
\bibitem[M{\"u}ller \emph{et~al.}(2016)M{\"u}ller, Drouin, Pearson, Ordu,
  Wehres, and Lewen]{Muller:2016de}
H.~S.~P. M{\"u}ller, B.~J. Drouin, J.~C. Pearson, M.~H. Ordu, N.~Wehres and
  F.~Lewen, \emph{Astronomy {\&} Astrophysics}, 2016, \textbf{586},
  A17--6\relax
\mciteBstWouldAddEndPuncttrue
\mciteSetBstMidEndSepPunct{\mcitedefaultmidpunct}
{\mcitedefaultendpunct}{\mcitedefaultseppunct}\relax
\EndOfBibitem
\bibitem[Carroll \emph{et~al.}(2010)Carroll, Drouin, and
  Widicus~Weaver]{Carroll:2010gt}
P.~B. Carroll, B.~J. Drouin and S.~L. Widicus~Weaver, \emph{The Astrophysical
  Journal}, 2010, \textbf{723}, 845--849\relax
\mciteBstWouldAddEndPuncttrue
\mciteSetBstMidEndSepPunct{\mcitedefaultmidpunct}
{\mcitedefaultendpunct}{\mcitedefaultseppunct}\relax
\EndOfBibitem
\bibitem[Medvedev and de~Lucia(2007)]{Medvedev:2007kl}
I.~R. Medvedev and F.~C. de~Lucia, \emph{ApJ}, 2007, \textbf{656},
  621--628\relax
\mciteBstWouldAddEndPuncttrue
\mciteSetBstMidEndSepPunct{\mcitedefaultmidpunct}
{\mcitedefaultendpunct}{\mcitedefaultseppunct}\relax
\EndOfBibitem
\bibitem[Saito \emph{et~al.}(1987)Saito, Kawaguchi, Yamamoto, Ohishi, Suzuki,
  and Kaifu]{Saito:1987fa}
S.~Saito, K.~Kawaguchi, S.~Yamamoto, M.~Ohishi, H.~Suzuki and N.~Kaifu,
  \emph{Astrophysical Journal}, 1987, \textbf{317}, L115--L118\relax
\mciteBstWouldAddEndPuncttrue
\mciteSetBstMidEndSepPunct{\mcitedefaultmidpunct}
{\mcitedefaultendpunct}{\mcitedefaultseppunct}\relax
\EndOfBibitem
\bibitem[Hirahara \emph{et~al.}(1993)Hirahara, Ohshima, and
  Endo]{Hirahara:1993ud}
Y.~Hirahara, Y.~Ohshima and Y.~Endo, \emph{Astrophysical Journal}, 1993,
  \textbf{408}, L113--L115\relax
\mciteBstWouldAddEndPuncttrue
\mciteSetBstMidEndSepPunct{\mcitedefaultmidpunct}
{\mcitedefaultendpunct}{\mcitedefaultseppunct}\relax
\EndOfBibitem
\bibitem[Yamamoto \emph{et~al.}(1987)Yamamoto, Saito, Kawaguchi, Kaifu, Suzuki,
  and Ohishi]{Yamamoto:1987jd}
S.~Yamamoto, S.~Saito, K.~Kawaguchi, N.~Kaifu, H.~Suzuki and M.~Ohishi,
  \emph{Astrophysical Journal}, 1987, \textbf{317}, L119--L121\relax
\mciteBstWouldAddEndPuncttrue
\mciteSetBstMidEndSepPunct{\mcitedefaultmidpunct}
{\mcitedefaultendpunct}{\mcitedefaultseppunct}\relax
\EndOfBibitem
\bibitem[Balle and Flygare(1981)]{Balle:1981ex}
T.~J. Balle and W.~H. Flygare, \emph{Review of Scientific Instruments}, 1981,
  \textbf{52}, 33--45\relax
\mciteBstWouldAddEndPuncttrue
\mciteSetBstMidEndSepPunct{\mcitedefaultmidpunct}
{\mcitedefaultendpunct}{\mcitedefaultseppunct}\relax
\EndOfBibitem
\bibitem[Oliveira \emph{et~al.}(2017)Oliveira, Martin-Drumel, and
  McCarthy]{jasmine:isms}
J.~Oliveira, M.-A. Martin-Drumel and M.~C. McCarthy, \emph{{SPECData: Automated
  Analysis Software for Broadband Spectra}}, 2017, For the current version, see
  https://github.com/jnicoleoliveira/SPECData\relax
\mciteBstWouldAddEndPuncttrue
\mciteSetBstMidEndSepPunct{\mcitedefaultmidpunct}
{\mcitedefaultendpunct}{\mcitedefaultseppunct}\relax
\EndOfBibitem
\bibitem[Martin-Drumel \emph{et~al.}(2016)Martin-Drumel, McCarthy, Patterson,
  McGuire, and Crabtree]{MartinDrumel:2016fw}
M.-A. Martin-Drumel, M.~C. McCarthy, D.~Patterson, B.~A. McGuire and K.~N.
  Crabtree, \emph{J Chem Phys}, 2016, \textbf{144}, 124202\relax
\mciteBstWouldAddEndPuncttrue
\mciteSetBstMidEndSepPunct{\mcitedefaultmidpunct}
{\mcitedefaultendpunct}{\mcitedefaultseppunct}\relax
\EndOfBibitem
\bibitem[Vrtilek \emph{et~al.}(1992)Vrtilek, Gottlieb, Gottlieb, Wang, and
  Thaddeus]{Vrtilek:1992jt}
J.~M. Vrtilek, C.~A. Gottlieb, E.~W. Gottlieb, W.~Wang and P.~Thaddeus,
  \emph{Astrophysical Journal}, 1992, \textbf{398}, L73--L76\relax
\mciteBstWouldAddEndPuncttrue
\mciteSetBstMidEndSepPunct{\mcitedefaultmidpunct}
{\mcitedefaultendpunct}{\mcitedefaultseppunct}\relax
\EndOfBibitem
\bibitem[{McCarthy} \emph{et~al.}(1994){McCarthy}, {Vrtilek}, {Gottlieb},
  {Tao}, {Gottlieb}, and {Thaddeus}]{mccarthy:l127}
M.~C. {McCarthy}, J.~M. {Vrtilek}, E.~W. {Gottlieb}, F.-M. {Tao}, C.~A.
  {Gottlieb} and P.~{Thaddeus}, \emph{The Astrophysical Journal Letters}, 1994,
  \textbf{431}, L127--L130\relax
\mciteBstWouldAddEndPuncttrue
\mciteSetBstMidEndSepPunct{\mcitedefaultmidpunct}
{\mcitedefaultendpunct}{\mcitedefaultseppunct}\relax
\EndOfBibitem
\bibitem[McCarthy \emph{et~al.}(1995)McCarthy, Gottlieb, Cooksy, and
  Thaddeus]{mccarthy:7779}
M.~C. McCarthy, C.~A. Gottlieb, A.~L. Cooksy and P.~Thaddeus, \emph{The Journal
  of Chemical Physics}, 1995, \textbf{103}, 7779--7787\relax
\mciteBstWouldAddEndPuncttrue
\mciteSetBstMidEndSepPunct{\mcitedefaultmidpunct}
{\mcitedefaultendpunct}{\mcitedefaultseppunct}\relax
\EndOfBibitem
\bibitem[Gottlieb \emph{et~al.}(2003)Gottlieb, Myers, and
  Thaddeus]{gottlieb:655}
C.~A. Gottlieb, P.~C. Myers and P.~Thaddeus, \emph{The Astrophysical Journal},
  2003, \textbf{588}, 655\relax
\mciteBstWouldAddEndPuncttrue
\mciteSetBstMidEndSepPunct{\mcitedefaultmidpunct}
{\mcitedefaultendpunct}{\mcitedefaultseppunct}\relax
\EndOfBibitem
\bibitem[Stanton \emph{et~al.}(2017)Stanton, Gauss, Cheng, Harding, Matthews,
  Szalay, Bartlett, Benedikt, Berger, Bernholdt, Bomble, Christiansen, Engel,
  Faber, Heckert, Heun, Huber, Jagau, Jonsson, Juselius, Klein, Lauderdale,
  Lipparini, Metzroth, Muck, O'Neill, Price, Prochnow, Puzzarini, Ruud,
  Schiffmann, Schwalbach, Simmons, Stopkowicz, Tajti, Vazquez, Wang, and
  Watts]{stanton_j_f_cfour_2017}
J.~F. Stanton, J.~Gauss, L.~Cheng, M.~E. Harding, D.~A. Matthews, A.~A.~A.
  Szalay, P~G, R.~J. Bartlett, U.~Benedikt, C.~Berger, D.~E. Bernholdt, Y.~J.
  Bomble, O.~Christiansen, F.~Engel, R.~Faber, M.~Heckert, O.~Heun, C.~Huber,
  T.~C. Jagau, D.~Jonsson, J.~Juselius, K.~Klein, W.~J. Lauderdale,
  F.~Lipparini, T.~Metzroth, L.~A. Muck, D.~P. O'Neill, D.~R. Price,
  E.~Prochnow, C.~Puzzarini, K.~Ruud, F.~Schiffmann, W.~Schwalbach, C.~Simmons,
  S.~Stopkowicz, A.~Tajti, J.~Vazquez, F.~Wang and J.~D. Watts, \emph{{CFOUR},
  {Coupled}-{Cluster} techniques for {Computational} {Chemistry}}, 2017, For
  the current version, see http://www.cfour.de.\relax
\mciteBstWouldAddEndPunctfalse
\mciteSetBstMidEndSepPunct{\mcitedefaultmidpunct}
{}{\mcitedefaultseppunct}\relax
\EndOfBibitem
\bibitem[Dunning(1989)]{DunningBasis}
T.~J. Dunning, \emph{The Journal of Chemical Physics}, 1989, \textbf{90},
  1007--1023\relax
\mciteBstWouldAddEndPuncttrue
\mciteSetBstMidEndSepPunct{\mcitedefaultmidpunct}
{\mcitedefaultendpunct}{\mcitedefaultseppunct}\relax
\EndOfBibitem
\bibitem[Harding \emph{et~al.}(2008)Harding, Vazquez, Ruscic, Wilson, Gauss,
  and Stanton]{harding_high-accuracy_2008}
M.~E. Harding, J.~Vazquez, B.~Ruscic, A.~K. Wilson, J.~Gauss and J.~F. Stanton,
  \emph{Journal of Chemical Physics}, 2008, \textbf{128}, 114111\relax
\mciteBstWouldAddEndPuncttrue
\mciteSetBstMidEndSepPunct{\mcitedefaultmidpunct}
{\mcitedefaultendpunct}{\mcitedefaultseppunct}\relax
\EndOfBibitem
\bibitem[Bomble \emph{et~al.}(2006)Bomble, Vazquez, Kallay, Michauk, Szalay,
  Csaszar, Gauss, and Stanton]{bomble_high-accuracy_2006}
Y.~J. Bomble, J.~Vazquez, M.~Kallay, C.~Michauk, P.~G. Szalay, A.~G. Csaszar,
  J.~Gauss and J.~F. Stanton, \emph{Journal of Chemical Physics}, 2006,
  \textbf{125}, 064108\relax
\mciteBstWouldAddEndPuncttrue
\mciteSetBstMidEndSepPunct{\mcitedefaultmidpunct}
{\mcitedefaultendpunct}{\mcitedefaultseppunct}\relax
\EndOfBibitem
\bibitem[K\'{a}llay \emph{et~al.}(2017)K\'{a}llay, Rolik, Csontos, Nagy, Samu,
  Mester, Cs\'{o}ka, Ladj\'{a}nszki, Szegedy, Lad\'{o}czki, Petrov, Farkas, and
  H\'{e}gely]{kallay_mrcc_2017}
M.~K\'{a}llay, Z.~Rolik, J.~Csontos, P.~Nagy, G.~Samu, D.~Mester, J.~Cs\'{o}ka,
  I.~Ladj\'{a}nszki, L.~Szegedy, B.~Lad\'{o}czki, K.~Petrov, M.~Farkas and
  B.~H\'{e}gely, \emph{{MRCC}, a quantum chemical program suite}, 2017, See
  www.mrcc.hu\relax
\mciteBstWouldAddEndPuncttrue
\mciteSetBstMidEndSepPunct{\mcitedefaultmidpunct}
{\mcitedefaultendpunct}{\mcitedefaultseppunct}\relax
\EndOfBibitem
\bibitem[Gottlieb \emph{et~al.}(1998)Gottlieb, McCarthy, Gordon, Chakan,
  Apponi, and Thaddeus]{Gottlieb:1998}
C.~A. Gottlieb, M.~C. McCarthy, V.~D. Gordon, J.~M. Chakan, A.~J. Apponi and
  P.~Thaddeus, \emph{The Astrophysical Journal Letters}, 1998, \textbf{509},
  L141\relax
\mciteBstWouldAddEndPuncttrue
\mciteSetBstMidEndSepPunct{\mcitedefaultmidpunct}
{\mcitedefaultendpunct}{\mcitedefaultseppunct}\relax
\EndOfBibitem
\bibitem[Tang and Saito(1995)]{Tang:1995jr}
J.~A. Tang and S.~Saito, \emph{J Mol Spectrosc}, 1995, \textbf{169},
  92--107\relax
\mciteBstWouldAddEndPuncttrue
\mciteSetBstMidEndSepPunct{\mcitedefaultmidpunct}
{\mcitedefaultendpunct}{\mcitedefaultseppunct}\relax
\EndOfBibitem
\bibitem[Dudek \emph{et~al.}(2017)Dudek, Salomon, Fangh{\"a}nel, and
  Thorwirth]{Dudek:2017ij}
J.~B. Dudek, T.~Salomon, S.~Fangh{\"a}nel and S.~Thorwirth, \emph{Int. J.
  Quantum Chem.}, 2017, \textbf{117}, e25414--12\relax
\mciteBstWouldAddEndPuncttrue
\mciteSetBstMidEndSepPunct{\mcitedefaultmidpunct}
{\mcitedefaultendpunct}{\mcitedefaultseppunct}\relax
\EndOfBibitem
\bibitem[Martin-Drumel \emph{et~al.}(2012)Martin-Drumel, Eliet, Pirali, Guinet,
  Hindle, Mouret, and Cuisset]{MartinDrumel:2012br}
M.~A. Martin-Drumel, S.~Eliet, O.~Pirali, M.~Guinet, F.~Hindle, G.~Mouret and
  A.~Cuisset, \emph{Chemical Physics Letters}, 2012, \textbf{550}, 8--14\relax
\mciteBstWouldAddEndPuncttrue
\mciteSetBstMidEndSepPunct{\mcitedefaultmidpunct}
{\mcitedefaultendpunct}{\mcitedefaultseppunct}\relax
\EndOfBibitem
\bibitem[Meerts and Dymanus(1975)]{Meerts:1975kq}
W.~L. Meerts and A.~Dymanus, \emph{Canadian Journal of Physics}, 1975,
  \textbf{53}, 2123--2141\relax
\mciteBstWouldAddEndPuncttrue
\mciteSetBstMidEndSepPunct{\mcitedefaultmidpunct}
{\mcitedefaultendpunct}{\mcitedefaultseppunct}\relax
\EndOfBibitem
\bibitem[Seeger \emph{et~al.}(1994)Seeger, Botschwina, Fl{\"u}gge, Reisenauer,
  and Maier]{Seeger:1994bp}
S.~Seeger, P.~Botschwina, J.~Fl{\"u}gge, H.~P. Reisenauer and G.~Maier,
  \emph{Journal of Molecular Structure: THEOCHEM}, 1994, \textbf{303},
  213--225\relax
\mciteBstWouldAddEndPuncttrue
\mciteSetBstMidEndSepPunct{\mcitedefaultmidpunct}
{\mcitedefaultendpunct}{\mcitedefaultseppunct}\relax
\EndOfBibitem
\bibitem[Thorwirth \emph{et~al.}(2017)Thorwirth, Salomon, Fangh{\"a}nel,
  Kozubal, and Dudek]{Thorwirth:2017fg}
S.~Thorwirth, T.~Salomon, S.~Fangh{\"a}nel, J.~R. Kozubal and J.~B. Dudek,
  \emph{Chemical Physics Letters}, 2017, \textbf{684}, 262--266\relax
\mciteBstWouldAddEndPuncttrue
\mciteSetBstMidEndSepPunct{\mcitedefaultmidpunct}
{\mcitedefaultendpunct}{\mcitedefaultseppunct}\relax
\EndOfBibitem
\bibitem[Pickett(1991)]{Pickett:1991cv}
H.~M. Pickett, \emph{J Mol Spectrosc}, 1991, \textbf{148}, 371--377\relax
\mciteBstWouldAddEndPuncttrue
\mciteSetBstMidEndSepPunct{\mcitedefaultmidpunct}
{\mcitedefaultendpunct}{\mcitedefaultseppunct}\relax
\EndOfBibitem
\bibitem[Zaleski \emph{et~al.}(2013)Zaleski, Seifert, Steber, Muckle, Loomis,
  Corby, Martinez, Crabtree, Jewell, Hollis, Lovas, Vasquez, Nyiramahirwe,
  Sciortino, Johnson, McCarthy, Remijan, and Pate]{Zaleski:2013bc}
D.~P. Zaleski, N.~A. Seifert, A.~L. Steber, M.~T. Muckle, R.~A. Loomis, J.~F.
  Corby, O.~J. Martinez, K.~N. Crabtree, P.~R. Jewell, J.~M. Hollis, F.~J.
  Lovas, D.~Vasquez, J.~Nyiramahirwe, N.~Sciortino, K.~Johnson, M.~C. McCarthy,
  A.~J. Remijan and B.~H. Pate, \emph{The Astrophysical Journal Letters}, 2013,
  \textbf{765}, L10\relax
\mciteBstWouldAddEndPuncttrue
\mciteSetBstMidEndSepPunct{\mcitedefaultmidpunct}
{\mcitedefaultendpunct}{\mcitedefaultseppunct}\relax
\EndOfBibitem
\bibitem[Lovas \emph{et~al.}(1980)Lovas, Suenram, Johnson, Clark, and
  Tiemann]{Lovas:1980jc}
F.~J. Lovas, R.~D. Suenram, D.~R. Johnson, F.~O. Clark and E.~Tiemann, \emph{J
  Chem Phys}, 1980, \textbf{72}, 4964--10\relax
\mciteBstWouldAddEndPuncttrue
\mciteSetBstMidEndSepPunct{\mcitedefaultmidpunct}
{\mcitedefaultendpunct}{\mcitedefaultseppunct}\relax
\EndOfBibitem
\bibitem[Brown \emph{et~al.}(1980)Brown, Godfry, and Winkler]{Brown:1980ga}
R.~D. Brown, P.~D. Godfry and D.~A. Winkler, \emph{Australian Journal of
  Chemistry}, 1980, \textbf{33}, 1\relax
\mciteBstWouldAddEndPuncttrue
\mciteSetBstMidEndSepPunct{\mcitedefaultmidpunct}
{\mcitedefaultendpunct}{\mcitedefaultseppunct}\relax
\EndOfBibitem
\bibitem[Takano \emph{et~al.}(1990)Takano, Sugie, Sugawara, Takeo, Matsumura,
  Masuda, and Kuchitsu]{Takano:1990hd}
S.~Takano, M.~Sugie, K.-i. Sugawara, H.~Takeo, C.~Matsumura, A.~Masuda and
  K.~Kuchitsu, \emph{J Mol Spectrosc}, 1990, \textbf{141}, 13--22\relax
\mciteBstWouldAddEndPuncttrue
\mciteSetBstMidEndSepPunct{\mcitedefaultmidpunct}
{\mcitedefaultendpunct}{\mcitedefaultseppunct}\relax
\EndOfBibitem
\bibitem[McCarthy \emph{et~al.}(2017)McCarthy, Zou, and
  Martin-Drumel]{mccarthy:154301}
M.~C. McCarthy, L.~Zou and M.-A. Martin-Drumel, \emph{The Journal of Chemical
  Physics}, 2017, \textbf{146}, 154301\relax
\mciteBstWouldAddEndPuncttrue
\mciteSetBstMidEndSepPunct{\mcitedefaultmidpunct}
{\mcitedefaultendpunct}{\mcitedefaultseppunct}\relax
\EndOfBibitem
\bibitem[Smith \emph{et~al.}(1988)Smith, Adams, Giles, and
  Herbst]{Smith:1988dc}
D.~Smith, N.~G. Adams, K.~Giles and E.~Herbst, \emph{Astronomy {\&}
  Astrophysics}, 1988, \textbf{200}, 191--194\relax
\mciteBstWouldAddEndPuncttrue
\mciteSetBstMidEndSepPunct{\mcitedefaultmidpunct}
{\mcitedefaultendpunct}{\mcitedefaultseppunct}\relax
\EndOfBibitem
\bibitem[Millar and Herbst(1990)]{Millar:1990yu}
T.~J. Millar and E.~Herbst, \emph{Astronomy {\&} Astrophysics}, 1990,
  \textbf{231}, 466--472\relax
\mciteBstWouldAddEndPuncttrue
\mciteSetBstMidEndSepPunct{\mcitedefaultmidpunct}
{\mcitedefaultendpunct}{\mcitedefaultseppunct}\relax
\EndOfBibitem
\bibitem[Cernicharo \emph{et~al.}(1987)Cernicharo, Kahane, Gu{\'e}lin, and
  Hein]{Cernicharo:1987jh}
J.~Cernicharo, C.~Kahane, M.~Gu{\'e}lin and H.~Hein, \emph{Astronomy {\&}
  Astrophysics}, 1987, \textbf{181}, L9--L12\relax
\mciteBstWouldAddEndPuncttrue
\mciteSetBstMidEndSepPunct{\mcitedefaultmidpunct}
{\mcitedefaultendpunct}{\mcitedefaultseppunct}\relax
\EndOfBibitem
\bibitem[Yamada \emph{et~al.}(2002)Yamada, Osamura, and Kaiser]{Yamada:2002gm}
M.~Yamada, Y.~Osamura and R.~I. Kaiser, \emph{Astronomy {\&} Astrophysics},
  2002, \textbf{395}, 1031--1044\relax
\mciteBstWouldAddEndPuncttrue
\mciteSetBstMidEndSepPunct{\mcitedefaultmidpunct}
{\mcitedefaultendpunct}{\mcitedefaultseppunct}\relax
\EndOfBibitem
\bibitem[Petrie(1996)]{Petrie:1996yd}
S.~Petrie, \emph{Monthly Notices of the Royal Astronomical Society}, 1996,
  \textbf{281}, 666--672\relax
\mciteBstWouldAddEndPuncttrue
\mciteSetBstMidEndSepPunct{\mcitedefaultmidpunct}
{\mcitedefaultendpunct}{\mcitedefaultseppunct}\relax
\EndOfBibitem
\bibitem[Sakai \emph{et~al.}(2007)Sakai, Ikeda, Morita, Sakai, Takano, Osamura,
  and Yamamoto]{Sakai:2007ud}
N.~Sakai, M.~Ikeda, M.~Morita, T.~Sakai, S.~Takano, Y.~Osamura and S.~Yamamoto,
  \emph{ApJ}, 2007, \textbf{663}, 1174--1179\relax
\mciteBstWouldAddEndPuncttrue
\mciteSetBstMidEndSepPunct{\mcitedefaultmidpunct}
{\mcitedefaultendpunct}{\mcitedefaultseppunct}\relax
\EndOfBibitem
\bibitem[Sakai \emph{et~al.}(2013)Sakai, Takano, Sakai, Shiba, Sumiyoshi, Endo,
  and Yamamoto]{sakai:9831}
N.~Sakai, S.~Takano, T.~Sakai, S.~Shiba, Y.~Sumiyoshi, Y.~Endo and S.~Yamamoto,
  \emph{The Journal of Physical Chemistry A}, 2013, \textbf{117},
  9831--9839\relax
\mciteBstWouldAddEndPuncttrue
\mciteSetBstMidEndSepPunct{\mcitedefaultmidpunct}
{\mcitedefaultendpunct}{\mcitedefaultseppunct}\relax
\EndOfBibitem
\bibitem[Seaver \emph{et~al.}(1982)Seaver, Hudgens, and Corpo]{seaver:63}
M.~Seaver, J.~W. Hudgens and J.~D. Corpo, \emph{Chemical Physics}, 1982,
  \textbf{70}, 63 -- 68\relax
\mciteBstWouldAddEndPuncttrue
\mciteSetBstMidEndSepPunct{\mcitedefaultmidpunct}
{\mcitedefaultendpunct}{\mcitedefaultseppunct}\relax
\EndOfBibitem
\bibitem[Lattanzi \emph{et~al.}(2010)Lattanzi, Gottlieb, Thaddeus, Thorwirth,
  and McCarthy]{lattanzi:1717}
V.~Lattanzi, C.~A. Gottlieb, P.~Thaddeus, S.~Thorwirth and M.~C. McCarthy,
  \emph{The Astrophysical Journal}, 2010, \textbf{720}, 1717\relax
\mciteBstWouldAddEndPuncttrue
\mciteSetBstMidEndSepPunct{\mcitedefaultmidpunct}
{\mcitedefaultendpunct}{\mcitedefaultseppunct}\relax
\EndOfBibitem
\bibitem[Ikeda \emph{et~al.}(1997)Ikeda, Sekimoto, and Yamamoto]{Ikeda:1997ty}
M.~Ikeda, Y.~Sekimoto and S.~Yamamoto, \emph{J Mol Spectrosc}, 1997,
  \textbf{185}, 21--25\relax
\mciteBstWouldAddEndPuncttrue
\mciteSetBstMidEndSepPunct{\mcitedefaultmidpunct}
{\mcitedefaultendpunct}{\mcitedefaultseppunct}\relax
\EndOfBibitem
\bibitem[Casavecchia \emph{et~al.}(2002)Casavecchia, Balucani, Cartechini,
  Capozza, Bergeat, and Volpi]{casavecchia:271}
P.~Casavecchia, N.~Balucani, L.~Cartechini, G.~Capozza, A.~Bergeat and G.~G.
  Volpi, \emph{Faraday Discuss.}, 2002, \textbf{119}, 27--49\relax
\mciteBstWouldAddEndPuncttrue
\mciteSetBstMidEndSepPunct{\mcitedefaultmidpunct}
{\mcitedefaultendpunct}{\mcitedefaultseppunct}\relax
\EndOfBibitem
\bibitem[McCarthy and Thaddeus(2005)]{mccarthy:174308}
M.~C. McCarthy and P.~Thaddeus, \emph{The Journal of Chemical Physics}, 2005,
  \textbf{122}, 174308\relax
\mciteBstWouldAddEndPuncttrue
\mciteSetBstMidEndSepPunct{\mcitedefaultmidpunct}
{\mcitedefaultendpunct}{\mcitedefaultseppunct}\relax
\EndOfBibitem
\bibitem[Sanz \emph{et~al.}(2003)Sanz, McCarthy, and Thaddeus]{Sanz:2003iya}
M.~E. Sanz, M.~C. McCarthy and P.~Thaddeus, \emph{J Chem Phys}, 2003,
  \textbf{119}, 11715--14\relax
\mciteBstWouldAddEndPuncttrue
\mciteSetBstMidEndSepPunct{\mcitedefaultmidpunct}
{\mcitedefaultendpunct}{\mcitedefaultseppunct}\relax
\EndOfBibitem
\bibitem[Sanz \emph{et~al.}(2005)Sanz, McCarthy, and Thaddeus]{Sanz:2005kc}
M.~E. Sanz, M.~C. McCarthy and P.~Thaddeus, \emph{J Chem Phys}, 2005,
  \textbf{122}, 194319--10\relax
\mciteBstWouldAddEndPuncttrue
\mciteSetBstMidEndSepPunct{\mcitedefaultmidpunct}
{\mcitedefaultendpunct}{\mcitedefaultseppunct}\relax
\EndOfBibitem
\bibitem[Cernicharo and Gu{\'e}lin(1996)]{Cernicharo:1996kd}
J.~Cernicharo and M.~Gu{\'e}lin, \emph{Astronomy {\&} Astrophysics}, 1996,
  \textbf{309}, L27\relax
\mciteBstWouldAddEndPuncttrue
\mciteSetBstMidEndSepPunct{\mcitedefaultmidpunct}
{\mcitedefaultendpunct}{\mcitedefaultseppunct}\relax
\EndOfBibitem
\bibitem[Remijan \emph{et~al.}(2007)Remijan, Hollis, Lovas, Cordiner, Millar,
  Markwick-Kemper, and Jewell]{Remijan:2007vp}
A.~J. Remijan, J.~M. Hollis, F.~J. Lovas, M.~A. Cordiner, T.~J. Millar, A.~J.
  Markwick-Kemper and P.~R. Jewell, \emph{The Astrophysical Journal}, 2007,
  \textbf{664}, L47\relax
\mciteBstWouldAddEndPuncttrue
\mciteSetBstMidEndSepPunct{\mcitedefaultmidpunct}
{\mcitedefaultendpunct}{\mcitedefaultseppunct}\relax
\EndOfBibitem
\bibitem[Br{\"u}nken \emph{et~al.}(2007)Br{\"u}nken, Gupta, Gottlieb, McCarthy,
  and Thaddeus]{Brunken:2007yd}
S.~Br{\"u}nken, H.~Gupta, C.~A. Gottlieb, M.~C. McCarthy and P.~Thaddeus,
  \emph{ApJ}, 2007, \textbf{664}, L43--L46\relax
\mciteBstWouldAddEndPuncttrue
\mciteSetBstMidEndSepPunct{\mcitedefaultmidpunct}
{\mcitedefaultendpunct}{\mcitedefaultseppunct}\relax
\EndOfBibitem
\bibitem[Broten \emph{et~al.}(1978)Broten, Oka, Avery, MacLeod, and
  Kroto]{Broten:1978iu}
N.~W. Broten, T.~Oka, L.~W. Avery, J.~M. MacLeod and H.~W. Kroto,
  \emph{Astrophysical Journal}, 1978, \textbf{223}, L105--L107\relax
\mciteBstWouldAddEndPuncttrue
\mciteSetBstMidEndSepPunct{\mcitedefaultmidpunct}
{\mcitedefaultendpunct}{\mcitedefaultseppunct}\relax
\EndOfBibitem
\bibitem[Loomis \emph{et~al.}(2016)Loomis, Shingledecker, Langston, McGuire,
  Dollhopf, Burkhardt, Corby, Booth, Carroll, Turner, and
  Remijan]{Loomis:2016jsa}
R.~A. Loomis, C.~N. Shingledecker, G.~Langston, B.~A. McGuire, N.~M. Dollhopf,
  A.~M. Burkhardt, J.~Corby, S.~T. Booth, P.~B. Carroll, B.~Turner and A.~J.
  Remijan, \emph{Monthly Notices of the Royal Astronomical Society}, 2016,
  \textbf{463}, 4175--4183\relax
\mciteBstWouldAddEndPuncttrue
\mciteSetBstMidEndSepPunct{\mcitedefaultmidpunct}
{\mcitedefaultendpunct}{\mcitedefaultseppunct}\relax
\EndOfBibitem
\bibitem[Matthews \emph{et~al.}(1984)Matthews, Irvine, Friberg, Brown, and
  Godfrey]{Matthews:1984hx}
H.~E. Matthews, W.~M. Irvine, P.~Friberg, R.~D. Brown and P.~D. Godfrey,
  \emph{Nature}, 1984, \textbf{310}, 125--126\relax
\mciteBstWouldAddEndPuncttrue
\mciteSetBstMidEndSepPunct{\mcitedefaultmidpunct}
{\mcitedefaultendpunct}{\mcitedefaultseppunct}\relax
\EndOfBibitem
\bibitem[McGuire \emph{et~al.}(2017)McGuire, Burkhardt, Shingledecker,
  Kalenskii, Herbst, Remijan, and McCarthy]{McGuire:2017ud}
B.~A. McGuire, A.~M. Burkhardt, C.~N. Shingledecker, S.~V. Kalenskii,
  E.~Herbst, A.~J. Remijan and M.~C. McCarthy, \emph{The Astrophysical Journal
  Letters}, 2017, \textbf{843}, L28\relax
\mciteBstWouldAddEndPuncttrue
\mciteSetBstMidEndSepPunct{\mcitedefaultmidpunct}
{\mcitedefaultendpunct}{\mcitedefaultseppunct}\relax
\EndOfBibitem
\bibitem[Kasai \emph{et~al.}(1993)Kasai, Obi, Ohshima, Hirahara, Endo,
  Kawaguchi, and Murakami]{Kasai:1993ds}
Y.~Kasai, K.~Obi, Y.~Ohshima, Y.~Hirahara, Y.~Endo, K.~Kawaguchi and
  A.~Murakami, \emph{Astrophysical Journal}, 1993, \textbf{410}, L45--L47\relax
\mciteBstWouldAddEndPuncttrue
\mciteSetBstMidEndSepPunct{\mcitedefaultmidpunct}
{\mcitedefaultendpunct}{\mcitedefaultseppunct}\relax
\EndOfBibitem
\bibitem[Ag{\'u}ndez \emph{et~al.}(2014)Ag{\'u}ndez, Cernicharo, and
  Gu{\'e}lin]{Agundez:2014gm}
M.~Ag{\'u}ndez, J.~Cernicharo and M.~Gu{\'e}lin, \emph{Astronomy {\&}
  Astrophysics}, 2014, \textbf{570}, A45--9\relax
\mciteBstWouldAddEndPuncttrue
\mciteSetBstMidEndSepPunct{\mcitedefaultmidpunct}
{\mcitedefaultendpunct}{\mcitedefaultseppunct}\relax
\EndOfBibitem
\bibitem[{Tieftrunk} \emph{et~al.}(1994){Tieftrunk}, {Pineau des Forets},
  {Schilke}, and {Walmsley}]{tieftrunk:579}
A.~{Tieftrunk}, G.~{Pineau des Forets}, P.~{Schilke} and C.~M. {Walmsley},
  \emph{Astronomy {\&} Astrophysics}, 1994, \textbf{289}, 579--596\relax
\mciteBstWouldAddEndPuncttrue
\mciteSetBstMidEndSepPunct{\mcitedefaultmidpunct}
{\mcitedefaultendpunct}{\mcitedefaultseppunct}\relax
\EndOfBibitem
\bibitem[Gupta \emph{et~al.}(2017)Gupta, Baraban, Changala, Thorwirth, Stanton,
  Martin-Drumel, Pirali, Gottlieb, and McCarthy]{ISMS}
H.~Gupta, J.~H. Baraban, B.~Changala, S.~Thorwirth, J.~F. Stanton, M.-A.
  Martin-Drumel, O.~Pirali, C.~A. Gottlieb and M.~C. McCarthy, International
  Symposium On Molecular Spectroscopy, 72nd Meeting, Abstract WF08\relax
\mciteBstWouldAddEndPuncttrue
\mciteSetBstMidEndSepPunct{\mcitedefaultmidpunct}
{\mcitedefaultendpunct}{\mcitedefaultseppunct}\relax
\EndOfBibitem
\bibitem[Yamamoto \emph{et~al.}(1990)Yamamoto, Saito, Kawaguchi, Chikada,
  Suzuki, Kaifu, Ishikawa, and Ohishi]{Yamamoto:1990wc}
S.~Yamamoto, S.~Saito, K.~Kawaguchi, Y.~Chikada, H.~Suzuki, N.~Kaifu,
  S.~Ishikawa and M.~Ohishi, \emph{The Astrophysical Journal}, 1990,
  \textbf{361}, 318--324\relax
\mciteBstWouldAddEndPuncttrue
\mciteSetBstMidEndSepPunct{\mcitedefaultmidpunct}
{\mcitedefaultendpunct}{\mcitedefaultseppunct}\relax
\EndOfBibitem
\bibitem[Sakai and Yamamoto(2013)]{Sakai:2013gg}
N.~Sakai and S.~Yamamoto, \emph{Chemical Reviews}, 2013, \textbf{113},
  8981--9015\relax
\mciteBstWouldAddEndPuncttrue
\mciteSetBstMidEndSepPunct{\mcitedefaultmidpunct}
{\mcitedefaultendpunct}{\mcitedefaultseppunct}\relax
\EndOfBibitem
\bibitem[Ohshima and Endo(1992)]{Ohshima:1992fa}
Y.~Ohshima and Y.~Endo, \emph{J Mol Spectrosc}, 1992, \textbf{153},
  627--634\relax
\mciteBstWouldAddEndPuncttrue
\mciteSetBstMidEndSepPunct{\mcitedefaultmidpunct}
{\mcitedefaultendpunct}{\mcitedefaultseppunct}\relax
\EndOfBibitem
\bibitem[Gordon \emph{et~al.}(2001)Gordon, McCarthy, Apponi, and
  Thaddeus]{Gordon:2001pd}
V.~D. Gordon, M.~C. McCarthy, A.~J. Apponi and P.~Thaddeus, \emph{ApJS}, 2001,
  \textbf{134}, 311--317\relax
\mciteBstWouldAddEndPuncttrue
\mciteSetBstMidEndSepPunct{\mcitedefaultmidpunct}
{\mcitedefaultendpunct}{\mcitedefaultseppunct}\relax
\EndOfBibitem
\bibitem[Ahrens and Winnewisser(1999)]{ahrens_pure_1999}
V.~Ahrens and G.~Winnewisser, \emph{Zeitschrift f\"{u}r Naturforschung}, 1999,
  \textbf{54a}, 131--136\relax
\mciteBstWouldAddEndPuncttrue
\mciteSetBstMidEndSepPunct{\mcitedefaultmidpunct}
{\mcitedefaultendpunct}{\mcitedefaultseppunct}\relax
\EndOfBibitem
\end{mcitethebibliography}
\bibliographystyle{rsc} %the RSC's .bst file

\newpage

%\begin{minipage}{\textwidth}

%%%%%%%%%%%%%%%%%%%%%%%%%%%%%%%%%%%%%%%%%%%%%%%%%%%%%%%%%%%%%%%%%%%%%%%%%%%%%%%%%%%%%%%%%
%%%%%%%%%%%%%%%%%%%%%%%%%%%  appendix                  %%%%%%%%%%%%%%%%%%%%%%%%%%%%%%%%%%
%%%%%%%%%%%%%%%%%%%%%%%%%%%%%%%%%%%%%%%%%%%%%%%%%%%%%%%%%%%%%%%%%%%%%%%%%%%%%%%%%%%%%%%%%

\appendix

\renewcommand{\thefigure}{A\arabic{figure}}
\renewcommand{\thetable}{A\arabic{table}}
\setcounter{figure}{0}
\setcounter{table}{0}

\begin{table*}[ht]
\centering
%\small
  \caption{\textit{Ab initio} vibrational frequencies and first order vibration-rotation coupling constants ($\alpha$) for ground state  \ce{C2S} ($\tilde{X}~^3\Sigma^-$) and  \ce{C4S} ($\tilde{X}~^3\Sigma^-$) computed at the fc-CCSD(T)/cc-pVDZ level of theory using VPT2. Frequencies are given in cm$^{-1}$ while $\alpha$ is provided in MHz.}
  \label{abinitio_data}
  \begin{tabular}{l c D{.}{.}{-1} D{.}{.}{-1} D{.}{.}{-1}}
    \toprule
    Species & Mode & \multicolumn{1}{c}{Harmonic frequency} &	\multicolumn{1}{c}{Anharmonic frequency} & \multicolumn{1}{c}{$\alpha$}	\\
    \midrule
    \ce{C2S} & $\nu_1$ & 1782.41 & 1632.77 & 46.62 \\
    & $\nu_3$ & 847.36 & 845.79 & 18.95 \\
   &  $\nu_2$ & 257.25 & 134.03 & -27.79 \\
    \midrule
   \ce{C4S} & $\nu_1$ & 2085.17 & 2043.99 & 6.82 \\
    & $\nu_2$ & 1783.96 & 1743.71 & 4.87 \\
    &$\nu_3$ & 1207.79 & 1208.14 & 3.52 \\
    &$\nu_4$ & 602.86  & 575.73  & 1.34 \\
    &$\nu_5$ & 519.47 & 290.92 & -1.50 \\
    &$\nu_6$ & 326.01 & 266.97 & -2.50 \\
    &$\nu_7$ & 129.15 & 112.42 & -3.21 \\
   \bottomrule
  \end{tabular}
\end{table*}

%\begin{figure*}[ht!]
%\includegraphics[width=1.\textwidth]{C4S_polyad.png}
%\includegraphics[width=0.49\textwidth]{C2S_vibE_annotated_v2.pdf}
%\includegraphics[width=0.49\textwidth]{C3S_vibE_annotated_v2.pdf}
%    \caption{\textcolor{red}{update for C4S only} Vibrational energy level diagram of \ce{C2S}, \ce{C3S}, and \ce{C4S} (left to right). The mode deformation associated with each fundamental vibration is pictorially represented. Vibrational quantum numbers for each state are indicated using the convention ($v_1\ v_2\ v_3 [\cdots]$) in which $v_i$ is the quanta of excitation in the $\nu_i$ modes. For brevity, the $l$ quantum number associated with the bending modes has been omitted. For \ce{C2S}, levels for which pure rotational transitions have been observed in this work are shown with plain lines (in red), while others are indicated with dashed lines (in gray).  Owing to the high density of states for \ce{C4S}, only observed and fundamental vibrational states are shown.}
%    \label{egy_diag_c4s}
%\end{figure*}

\begin{table*}[ht!]
\centering
  %\scriptsize
  \caption{Oxygen-, nitrogen-, sulfur-, and argon-containing species identified in each reaction mixture whose frequencies were known prior to present work. Numbers  in the Table represent the signal-to-noise ratio of the strongest line assigned for each species.}
  \label{known_molecules_on}
  \begin{tabular}{l cccc} 
    \toprule
   Species$^a$ 			&	\ce{CS2 + HC4H}	&	 \ce{CS2 + HCCH}	&	 \ce{CS2}			&	\ce{HC4H}		\\
    \midrule	
	
	\ce{C3O}       			            &   	& 5  	&       & 21 	\\
	\ce{C5O}   			                &   	&   	&   	&  3 	\\ 
	
	\vspace{-0.25em} \\
	\midrule[0.25pt]
	
	\ce{HC3O}     			            & 13	& 2 	&   	& 9  	\\ 
	\ce{HC4O}     			            &  7 	&   	&   	& 15  	\\ 
	\ce{HC5O}      			            &    	&   	&   	& 3  	\\ 
	\ce{HC6O}     			            &    	&   	&   	& 3  	\\ 
	\ce{HC7O}       		 	        &    	&   	&   	& 3  	\\ 

	\vspace{-0.25em} \\
	\midrule[0.25pt]

	\ce{H2CO}       			        &  4  	&   	&   	& 8  	\\
	\ce{H2C5O}      			        &  7  	&   	&   	& 8  	\\ 
	
	\vspace{-0.25em} \\
	\midrule[0.25pt]

	\ce{CH3CHO}   		                &   	& 99	&   	&  		\\ 
	\ce{CH3OCHO}			            &   	& 71	&   	&  		\\ 
	\ce{HCOC2H}   	                	& 6 	& 5 	&   	& 7 	\\ 
	\ce{C2H5OH}   			            &   	& 2 	&   	&  		\\ 
	OH                   			    &   	&  2    &   	&  		\\

	\vspace{-0.25em} \\
	\midrule[0.25pt]
	
	\ce{SO2}	 		            	& 2	    & 163	&		& 		\\
	\hspace{1em} $v_1=1$				&	    & 17	&		&		\\
	\hspace{1em} $v_2=1$				&		& 77	&   	&		\\
	%\hspace{1em} $v_1=1$				&	    & 17	&		&		\\ %overlapped with another line
    \hspace{1em} $v_2=2$			    &		& 29	&		&		\\
	\hspace{1em} $v_2=3$				&		& 3		&		&		\\
	\hspace{1em} $(v_1,v_2)=(1,1)$		&		& 12	&		&		\\

	\vspace{-0.5em} \\
	
	\hspace{1em} \ce{^{34}SO2}			& 		& 11	&		&		\\
	%\hspace{1em} \ce{^{34}SO2} $v_2=1$ 	&		& X 	&		&		\\ %Frequencies not found in the literature

	\vspace{-0.25em} \\
	\midrule[0.25pt]
	 
	\ce{HC3N}			                &  		& 4 	&		& 3		\\
	\ce{HC5N}      			            & 9		&		&		& 15	\\
	\ce{HC7N}  	    		            &  		&		&		& 3		\\ 

	\vspace{-0.25em} \\
	\midrule[0.25pt]

	\ce{CH3CN}         		            &  		& 3  	&   	&    	\\ 
	\ce{CH3NO}                 	        &   	& 3  	&   	&    	\\ 

 	\vspace{-0.25em} \\
	\midrule[0.25pt]
 
 	%Ar--\ce{HC4H}	  		            &   	&		&		&		\\  %%Not found anymore, must have been a coincidence
	\ce{H2O}--\ce{H2O}		            &  3	& 2		& 5		&		\\
	\ce{H2O}--\ce{HC2H}       	        &   	& 3  	&		&		\\ 
	\ce{H2O}--\ce{HC4H}       	        & 11	&   	&		&		\\ 

	\vspace{-0.25em} \\
	\midrule[0.25pt]
	
	OCS					                & 5 	& 6 	& 2		&		\\ 	
	SNO               			        &   	& 3  	&   	&    	\\ 	
    \bottomrule
  \end{tabular}
  
  \smallskip
  \begin{minipage}{0.5\textwidth}
  $^a$ Main isotopologue in its ground vibrational state, unless otherwise noted.
  \end{minipage}
\end{table*}

\begin{table*}
\centering
  \caption{Measured centimeter-wave transitions of the ground and vibrationally excited normal and isotopic \ce{C2S} (in MHz).}
    \begin{tabular}{ll D{.}{.}{-1} D{.}{.}{-1} D{.}{.}{-1} D{.}{.}{-1} D{.}{.}{-1} D{.}{.}{-1} D{.}{.}{-1} }
    \toprule
    	%\vspace{-0.75em}\\
 Iso. &  Vib.  &   \multicolumn{7}{c}{Transition, $J'_{N'} - J''_{N''}$}\\ 
\cmidrule{3-9}
Species &   State    & \multicolumn{1}{c}{$ 2_1 - 1_0$} &	\multicolumn{1}{c}{$1_2 - 2_1$} & \multicolumn{1}{c}{$2_3 - 1_2$}  & \multicolumn{1}{c}{$3_4 - 2_3$} & \multicolumn{1}{c}{$2_2 - 1_1$} & \multicolumn{1}{c}{$3_2 - 0_1$} & \multicolumn{1}{c}{$3_3 - 2_2$} \\
    \midrule
	%\vspace{-0.75em}\\
C$_2^{\,34}$S    &  GS $^a$          &         	&	21930.4756	&	33111.8370	&   &&&	\\
\ce{C2S}         &$(v_1,v_3)=(1,1)$ &    11027.0403	&	22154.5032	&	33457.0705  &	44971.725 &&&	\\	 	
\ce{C2S}         &   $v_3=3$  	   & 	11053.8201	&	22205.5507	&	33528.4000 	&	45059.175 &&&	\\	 	
\ce{C2S}         &   $v_1=1$  	   & 	11036.2863	&	22177.3620	&	33500.5144	&	45043.037 &&&	\\	 	
\ce{C2S}         &   $v_3=2$        &	11078.3628  &	22256.5671  &	33609.0152  & &&&	             \\
\ce{C2S}         &   $v_3=1$        &	11102.8764  &	22307.6307  &	33689.9173  &	45287.000  &&&   \\
\ce{C2S}         &$(v_1,v_2)=(1,1^1)$ &    -	&	22125.9220	&	33452.0276  &	45019.275 &&&	\\	 
            &                  &    -    &   22179.3904  &   33530.1076  &   45120.400 &&& \\
\ce{C2S}         &    GS $^a$        &    11119.4452	&	22344.0305	&	33751.3699	&  45379.020 & 25911.015 & 29477.700 & 38866.417	\\
\ce{C2S}         &$(v_3,v_2)=(1,1^1)$&    -	&   22270.2587  &   33658.5184  &   45280.525  &&&  \\
            &                  &    -	&   22323.5767  &   33736.5616  &   45381.850 &&&  \\
\ce{C2S}         &    $v_2=1^1$     &    -	&   22304.5697  &   33718.0178  &   45371.475  &&&  \\
            &                  &    -	&   22357.3069  &   33795.1252  &   45471.425  &&& \\
\ce{C2S}         &$(v_2,v_3)=(2^0,1)$&	11158.0445  &	22425.5719  &	33882.7887  &	45567.600   &&& \\
\ce{C2S}         &    $v_2=2^0$     &	11174.1680  &	22461.5174  &	33944.4177  &	45661.000   &&& \\
%% need to add C13CCS data
    \bottomrule
     \end{tabular}
     
    \smallskip
    \begin{minipage}{0.9\textwidth}
        Note: Estimated measurement uncertainties are 2\,kHz below 40\,GHz. Above this frequency, transitions have been measured using double resonance techniques resulting in a 25\,kHz uncertainty. For states with $l=1$, the $2_1 - 1_0$ transition is not allowed, as indicated by a dash symbol in the corresponding lines.\\
        $^a$ Several centimeter-wave lines were previously reported with a higher uncertainty in Refs.~\citenum{Saito:1987fa} and \citenum{Yamamoto:1990wc}.
    \end{minipage}
  \label{c2s_cm_freqs} 
\end{table*}

\begin{table*} \centering
\caption{Measured centimeter--wave transitions of $^{13}$C isotopologues of \ce{C2S} (in MHz).}
\begin{tabular}{D{-}{~-~}{-1} D{-}{~-~}{-1} D{.}{.}{-1} D{.}{.}{-1}}
    \toprule
    N'_{J'}-N''_{J''} & F' - F'' & \multicolumn{1}{c}{$^{13}$CCS} & \multicolumn{1}{c}{C$^{13}$CS} \\ \midrule
 2_1 - 1_0  &   0.5 - 0.5 &  10699.9518 & 11078.4014 \\
            &   1.5 - 0.5 &  10706.9119 & 11075.2940 \\
 1_2 - 2_1  &   1.5 - 1.5 &  21487.4534 & 22259.7156 \\
            &   1.5 - 0.5 &  21494.4117 & 22256.6071 \\
            &   2.5 - 1.5 &  21498.6616 & 22254.7319 \\ 
 2_3 - 1_2  &   2.5 - 1.5 &  32440.1924 & 33615.1718 \\
            &   3.5 - 2.5 &  32443.9526 & 33613.5400 \\
    \bottomrule
\end{tabular}
     
    \smallskip
    \begin{minipage}{0.9\textwidth}
        Note: Estimated measurement uncertainties are 2\,kHz.  Several centimeter-wave lines were previously reported with a higher uncertainty in Ref.~\citenum{Ikeda:1997ty}.
    \end{minipage}
  \label{c2s_c13_freqs} 
\end{table*}

\begin{table*}
    \centering
  \caption{Measured millimeter-wave transitions in the ground, $v_1=1$ and $v_3=1$ vibrationally excited states of  \ce{C2S} (in MHz).}
  \begin{tabular}{D{-}{~-~}{-1} l c c}
    \toprule
    	J'_{N'} - J''_{N''}	    &	\multicolumn{1}{c}{GS$^a$}  & $v_1=1$	&	$v_3=1$	\\
    \midrule
    0_{1} - 1_{2}    &	 162749.178      &             &                    \\
	1_{1} - 2_{1}    &	 183257.261      & 181634.082  & 184483.054		\\
	2_{2} - 1_{2}	 &	 186824.217      &             & 					\\
	0_{1} - 1_{0}    &	 196212.630      &             & 					\\
	3_{2} - 2_{1}    &	 214570.887      &             & 					\\
    20_{19} - 19_{18}&	258274.283       & 256395.149  &	257460.718    \\
    20_{20} - 19_{19}&	259055.427 	     &             &	258254.193 	\\
	20_{21} - 19_{20}&	259700.932       & 257811.769  &	258910.073     \\
	21_{20} - 20_{19}&	271292.242       & 269318.144  &	270439.575 		\\
    21_{21} - 20_{20}&	272002.244       &             &	271160.971 		\\
	21_{22} - 20_{21}&	272592.955       & 270610.151  &	271761.284 		\\
    \bottomrule
     \end{tabular}

  \smallskip
  \begin{minipage}{0.9\textwidth}
        Note: Estimated measurement uncertainties are 25\,kHz. Transitions below 250 GHz have been measured using double resonance technique in the cavity FT instrument.\\
        $^a$ Some of these transitions were previously reported by Refs.~\citenum{Saito:1987fa} and \citenum{Yamamoto:1990wc} with a 20 kHz uncertainty.%\\
        %\textcolor{red}{Add a comment about not seing the $v_2=1$ (and more) satellites (likely a problem of extrapolation from the fit)?}
  \end{minipage}
  \label{c2s_mm_freqs} 
\end{table*}

\clearpage

\begin{sidewaystable*}[p]
\centering
%\small
\caption{Spectroscopic constants of the ground and vibrationally excited normal and isotopic \ce{C2S} (in MHz, sorted by increasing $B$ value or by isotopic variant).}
\label{c2s_constants}
  \begin{tabular}{ll l l l l l l l l c l }
    \toprule
  Iso. &     	Vib. 		         &   	$B$	 &   $10^3D$  & 	$\gamma$  &  $10^3\gamma_D$ & 	$\lambda$  &  	$10^3\lambda_D$	 &	$q/2$      &  	$p/2$ & 	N$^a$   & 	weighted  \\
 Species &      	 State		         &    	 &     &    &    & 	   &   	 &      &   &  	  &  ave.$^b$  \\
    \midrule	 
$^{13}$CCS      & GS $^c$         &	6188.0867(4)    & 	1.5720(5)   &	-14.06(1)   &   0.037 &  97204.0(2)    & 24.5(3)&      &      &  56 &  0.83 \\
C$_2^{\;34}$S   & GS              &	6335.8839(3)   	& 	1.6543(5)		&	-14.386(7)		&  0.037	     &  97195.1(1)    & 26.8(2)   &       &      &  30 &  0.99 \\
\ce{C2S}        &$(v_1,v_3)=(1,1)$  &	6411.057(4)   	& 	1.7271		&	-14.711		&  0.037	     &  97323.7(3)    & 27.0   &      &      &  4  &  0.22 \\
\ce{C2S}        &$v_3=3$    & 6417.230(4)     &	1.7271		&	-14.711      &  0.037	     &	98259.4(4)    & 27.0   &      &      &  4 & 0.62\\
\ce{C2S}        & $v_1=1$	        & 6430.6293(3)	&	1.7271		&	-12.353(3) 	&  0.037	     &  96341.31(1)    & 27.0   &      &      &  9  &  0.57\\
\ce{C2S}        &$v_3=2$    & 6437.336(4)     &	1.7271		&	-14.711      &  0.037	     &	98039.0(4)    & 27.0   &      &     &  3  & 0.19\\
C$^{13}$CS      &GS $^c$            &	6446.9655(5)    & 	1.7119(7)   &	-14.63(1)   &  0.037 &  97226.7(2)    & 28.1(4)&      &     & 45 &  0.79 \\
\ce{C2S}        &$v_3=1$    & 6457.7175(2)    &	1.7271		&	-14.542(3)   &  0.037	     &	97800.18(1)    & 27.0   &      &      &  11  & 1.11\\
\ce{C2S}        &$(v_1,v_2)=(1,1^1)$&    6460.987(3)     &	1.7271		&	-14.711      &  0.037	     &	94920.1(2)    & 27.0   &  -4.597(5)&  -26.47(3)&  6 &4.51\\
\ce{C2S}    & GS      & 6477.7496(2)    &	1.7271(3)   &	-14.711(2)	&  0.037(5)	 &	97195.651(6)  & 27.0(2)&      &       &  52 & 0.78  \\ 
\ce{C2S}        &$(v_2,v_3)=(1^1,1)$&    6487.104(3)     &	1.7271		&	-14.711      &  0.037	     &	96699.5(3)    & 27.0   & -4.716(5)&  -25.79(3) &  6  & 3.67\\
\ce{C2S}        &$v_2=1^1$  & 6507.496(3)     &	1.7271		&	-14.711      &  0.037	     &	96077.5(3)    & 27.0     & -4.609(5) &  -25.72(3)&  6  & 3.58\\
\ce{C2S}        &$(v_2,v_3)=(2^0,1)$&    6513.254(4)	    &	1.7271		&	-14.711	    &  0.037	     &	96614.2(3)	  & 27.0   &      &     &  4  & 0.54 \\
\ce{C2S}        &$v_2=2^0$  & 6534.040(4)	    &	1.7271		&	-14.712	    &  0.037	     &	95972.5(3)	  & 27.0   &      &      &  4  & 0.35 \\
    \bottomrule
    \\
    \toprule
\ce{C2S}    & GS      & 6477.7496(2)    &	1.7271(3)   &	-14.711(2)	&  0.037(5)	 &	97195.651(6)  & 27.0(2)&      &       &  52 & 0.78  \\ 
            & $v_1=1$   & 6430.6293(3)	&	1.7271		&	-12.353(3) 	&  0.037	     &  96341.31(1)    & 27.0   &      &      &  9  &  0.57\\
            &$v_2=1^1$  & 6507.496(3)     &	1.7271		&	-14.711      &  0.037	     &	96077.5(3)    & 27.0     & -4.609(5) &  -25.72(3)&  6  & 3.58\\
            &$v_2=2^0$  & 6534.040(4)	    &	1.7271		&	-14.712	    &  0.037	     &	95972.5(3)	  & 27.0   &      &      &  4  & 0.35 \\
            &$v_3=1$    & 6457.7175(2)    &	1.7271		&	-14.542(3)   &  0.037	     &	97800.18(1)    & 27.0   &      &      &  11  & 1.11\\
            &$v_3=2$    & 6437.336(4)     &	1.7271		&	-14.711      &  0.037	     &	98039.0(4)    & 27.0   &      &     &  3  & 0.19\\
            &$v_3=3$    & 6417.230(4)     &	1.7271		&	-14.711      &  0.037	     &	98259.4(4)    & 27.0   &      &      &  4 & 0.62\\
            &$(v_1,v_2)=(1,1^1)$&    6460.987(3)     &	1.7271		&	-14.711      &  0.037	     &	94920.1(2)    & 27.0   &  -4.597(5)&  -26.47(3)&  6 &4.51\\
            &$(v_1,v_3)=(1,1)$  &	6411.057(4)   	& 	1.7271		&	-14.711		&  0.037	     &  97323.7(3)    & 27.0   &      &      &  4  &  0.22 \\
            &$(v_2,v_3)=(1^1,1)$&    6487.104(3)     &	1.7271		&	-14.711      &  0.037	     &	96699.5(3)    & 27.0   & -4.716(5)&  -25.79(3) &  6  & 3.67\\
            &$(v_2,v_3)=(2^0,1)$&    6513.254(4)	    &	1.7271		&	-14.711	    &  0.037	     &	96614.2(3)	  & 27.0   &      &     &  4  & 0.54 \\
C$_2^{\;34}$S   & GS              &	6335.8839(3)   	& 	1.6543(5)		&	-14.386(7)		&  0.037	     &  97195.1(1)    & 26.8(2)   &       &      &  30 &  0.99 \\
$^{13}$CCS      & GS $^c$         &	6188.0867(4)    & 	1.5720(5)   &	-14.06(1)   &   0.037 &  97204.0(2)    & 24.5(3)&      &      &  56 &  0.83 \\
C$^{13}$CS      &GS $^c$            &	6446.9655(5)    & 	1.7119(7)   &	-14.63(1)   &  0.037 &  97226.7(2)    & 28.1(4)&      &     & 45 &  0.79 \\
\bottomrule    
   \end{tabular}
   
   \smallskip
  \begin{minipage}{0.94\textwidth}
  Note: Uncertainties (1$\sigma$) are in units of the last significant digit. Best-fit constants derived from pure rotational frequencies reported in the literature\cite{Saito:1987fa, Yamamoto:1990wc, Ikeda:1997ty} and line frequencies in Tables~\ref{c2s_cm_freqs}, \ref{c2s_c13_freqs}, and \ref{c2s_mm_freqs}, using a standard linear molecule Hamiltonian in a $^3\Sigma$ electronic state, with or without $l$-type doubling. Values with no associated uncertainties were constrained to the value derived for the normal isotopic species. We note that the RMS values involving the $\nu_2$ state are significantly larger than those for other vibrational states. This difference arises in part due to the small dataset combined with the need to include several lambda-doubling terms.  A smaller RMS should be achieved by varying additional terms, but, for simplicity, we have chosen to report a fit in which only the leading constants were varied, and are well determined.\\
  $^a$ Refers to the number of lines in the fit.\\
  $^b$ Dimensionless.\\
  $^c$ The $^{13}$C hyperfine terms $b$ and $c$ are omitted here and are reported in Table \ref{tab:c2s_hfs_cts}
  \end{minipage}
\end{sidewaystable*}

\begin{table*}[ht]
    \caption{Hyperfine spectroscopic constants of $^{13}$C isotopic variants of \ce{C2S} (in MHz),}
    \centering
    \begin{tabular}{c D{.}{.}{-1} D{.}{.}{-1}}
        \toprule
        Parameter & \multicolumn{1}{c}{$^{13}$CCS} & \multicolumn{1}{c}{C$^{13}$CS} \\ \midrule
        $b$ & 18.6(6) & -19.2(6)\\
        $c$ & -50.(2) & -16.(2)\\
         \bottomrule
    \end{tabular}
    \label{tab:c2s_hfs_cts}
\end{table*}

%----------------------------------------------------------------------------------------------
% C3S
%----------------------------------------------------------------------------------------------
\clearpage

\begin{table*}[t]
\centering
  \caption{Measured centimeter-wave transitions of the ground and vibrationally excited normal and isotopic \ce{C3S} (in MHz).}
  
  \begin{tabular}{@{\extracolsep{\fill}}l l lllll}
    \toprule
Iso.  & Vib. &  \multicolumn{5}{c}{Transition, $J' - J''$}  \\ 
\cmidrule{3-7}
Species & State  & \multicolumn{1}{c}{$2 - 1$} &	\multicolumn{1}{c}{$3 -2$} & \multicolumn{1}{c}{$4 - 3$}  & \multicolumn{1}{c}{$5 - 4$} & \multicolumn{1}{c}{$6- 5$} \\
    \midrule
	\vspace{-0.75em}\\
%\cline{1}	          
$^{13}$CCCS  	&	$v_3=1$  	& 	11117.8715  &	16676.7930	&	22235.7098	 &  27794.5918   & 33353.4555  \\	 	
$^{13}$CCCS  	&	GS  	    & 	11132.2395$^a$  &	16698.3481$^a$	&   22264.4418$^a$	 &  27830.5140$^a$   & 33396.5590  \\ 
C$_3^{\,34}$S  	&	$v_3=1$  	& 	11266.5956  &	16899.8806	&	22533.1503	 &  28166.3989   & 33799.6182  \\
C$_3^{\,34}$S  	&	GS       	& 	11281.468$^b$  &	16922.1924$^a$	&	22562.8997$^a$	 &  28203.5870$^a$   & 33844.2426  \\ 
C$_3^{\,34}$S  	&	$v_4=1^1$  	&	11300.6585  &	16950.9732  &	22601.2722	 &  28251.5511   & 33901.8087  \\
			  	&             	& 	11306.4008  & 	16959.5927  & 	22612.7667	 &  28265.9189   & 33919.0458  \\
C$^{13}$CCS		& 	$v_3=1$  	&   11430.3317  &   17145.4856	&   22860.6204	 &  28575.7375   & 34290.8264  \\
C$^{13}$CCS  	&	GS  	    & 	11445.4767$^a$  &	17168.2027$^a$	&   22890.9123$^a$	 &  28613.6016$^a$   & 34336.2592  \\ 
C$_3$S		    &$(v_1,v_3)=(1,2)$& 11470.6766  &   17206.0013  &   22941.3100   &  28676.5972   & 34411.8556  \\ 
C$_3$S		    &$(v_1,v_3)=(1,1)\, ^{d,e}$&11486.9578 &   17230.4235  &   22973.8700   &  28717.2944   &             \\ 
                &                 & 11487.0135  &   17230.5062  &   22973.9831   &  28717.4367   &              \\
C$_3$S          & $v_3=4$         & 11500.9602  &   17251.4245  &   23001.8723   &  28752.2983   &             \\
C$_3$S			&$(v_2,v_3)=(1,1)$&	11501.4629	&   17252.1815  &   23002.8831   &  28753.5630   &  34504.2175 \\
C$_3$S          &  $v_1=1\, ^e$   &	11502.2624  &	17253.3805  &	23004.4807   &	28755.5607   &  34506.6142 \\
C$_3$S			&   $v_3=3$       &	11515.8557	&   17273.7704	&   23031.6688   &  28789.5478   &  34547.3957 \\
C$_3$S			&   $v_2=1$       &	11516.7518	&   17275.1133	&   23033.4615   &  28791.7862   &             \\
C$_3$S			&   $v_3=2$       &	11530.9586	&   17296.4249	&   23061.8746   &  28827.3029   &  34592.7057 \\
C$_3$S          &$(v_2,v_4)=(1,1^1)$& 11535.7961  &   17303.6763 	& 	23071.5442   &  28839.3851   &  34607.2016\\
                &				  & 11541.8122  & 	17312.7127	& 	23083.5936   &  28854.4462   &  34625.2725 \\
C$_3$S			&   $v_3=1$       &	11546.1972$^c$	&	17319.283$^c$  &   23092.3522$^c$   &  28865.4002$^c$  &  34638.4213$^c$ \\	
C$_3$S			&   GS            &	11561.5099$^a$	&	17342.2564$^a$  &   23122.9836$^a$   &  28903.6913$^a$   &  34684.3676 \\ 	
C$_3$S          &$(v_3,v_4)=(1,1^1)$&   11568.3746  &   17352.5444 	& 	23136.7006   &  28920.8350   &  34704.9434 \\
                &				  & 	11574.9783  & 	17362.4574 	& 	23149.9186   &  28937.3568   &  34724.7708 \\
C$_3$S			&   $v_4=1^1\, ^f$  &	11581.0824	&	17371.6094  &	23162.1189	 &	28952.6085   &  34743.0697 \\		
                &                 &	11587.1082  &	17380.6511  &	23174.1755	 &	28967.6795 	 &	34761.1548 \\ 
C$_3$S			&   $v_5=1^1\, ^f$  &   11602.8885	&	17404.3150	&   23205.7261	 &  29007.1130	 &  34808.4727 \\
                &                 &	11618.7422	&	17428.1009  &	23237.4398	 &	29046.7555	 &	34856.0407 \\
C$_3$S			&   $v_4=2^0$     &   11603.1367	&   17404.6924  &   23206.2315   &  29007.7498   &  34809.2416 \\
C$_3$S			&   $v_4=2^2$     &   \multicolumn{1}{c}{--}	&   17409.7432  &   23212.9647   &  29016.1635   &  34819.3362 \\
%C$_3$S			&$\nu_4+\nu_5^0$&   11619.9023	&   17429.8318  &   23239.7364   &  29049.6051  \\  %Mike:can someone else review this assignment?  This are real series!
%C$_3$S			&$\nu_4+\nu_5^2$&   $\cdots$	&   17429.9403  &   23239.8926   &  29049.8253   \\
C$_3$S			&   $v_5=2^{0\, ^f}$  &   11660.5482	&   17490.7982  &   23321.0138   &  29151.1892   &             \\
    \bottomrule
  \end{tabular}
  
  \smallskip
  \begin{minipage}{0.9\textwidth}
  Note: This work, unless otherwise noted. Estimated measurement uncertainties are 2\,kHz.  Previously identified isotopic species and vibrationally excited states are included for completeness.\\
  $^a$ Ref. \citenum{Sakai:2013gg}.\\
  $^b$ Ref. \citenum{Ohshima:1992fa}.\\
  $^c$ Ref. \citenum{Crabtree:2016fj}.\\
  $^d$ A closely-spaced doublet was observed.  The centroid was used in the least-squares fit.\\
  $^e$ First observation of the centimeter-wave transitions; assignments based on the infrared measurements in Ref.~\citenum{Dudek:2017ij}.\\
  $^f$ First observation of the centimeter-wave transitions; assignments based on the millimeter observations in Ref.~\citenum{Tang:1995jr}.
  \end{minipage}
  \label{c3s_cm_freqs}
\end{table*}

\newpage

\begin{sidewaystable*}[ht]
\centering
  \captionof{table}{\ Measured millimeter-wave transitions of \ce{C3S} (in MHz).}
  
  \begin{tabular}{D{-}{~-~}{-1}  c c c  c c c c c c }
    \toprule
J' - J''&   GS$^a$      &	$v_3=1$     & $v_3=2$       & $v_4=1^a$     & $v_4=2^0$	    &$v_4=2^{2,~b}$ & $v_5=1^a$ & $v_5=2^{0\, a}$&$(v_3,v_4)=(1,1^1)$	\\
    \midrule
44 - 43 &   254277.025  &   253940.217  &   253605.042  & 254706.569    &  255194.240   &   255261.620  & 255181.712    &  256384.431  &               \\
	    &               &               &               & 254839.022    &               &   255264.687  & 255526.635    &              &   254573.768  \\
45 - 44	&   260052.472  &   259708.080  &   259365.169  & 260491.715    &  260990.594   &   261059.157  & 260977.450    &  262204.508  &   260206.900  \\
	    &               &               &               & 260627.185    &               &   $\cdots$    & 261330.010    &              &               \\
46 - 45 &   265827.668  &   265475.613  &               & 266276.624    &  266786.708   &   266856.464  & 266772.916    &  268024.150  &               \\
	    &               &	            &	         	& 266415.084    &               &   266860.017  & 267133.140    &              &   266137.984  \\
47 - 46	&   271602.622  &	271242.975  &               & 272061.270    &  272582.580   &   272653.499  & 272568.145    &  273843.340  &               \\
	    &               &	            &	         	& 272202.742    &               &   272657.295  & 272935.981    &              &   271919.853  \\
48 - 47	&   277377.316  &	277009.990  &   276644.223  & 277845.724    &  278378.196   &   278450.302  &               &               &               \\
	    &               &	            &	         	& 277990.158    &               &   278454.324  &               &               &               \\
    \bottomrule
  \end{tabular}

\bigskip
  \begin{tabular}{D{-}{~-~}{-1}  c c c }
    \toprule
J' - J''& C$_3\: ^{34}$S$^a$& $^{13}$CCCS & C$^{13}$CCS	\\
    \midrule
44 - 43 &   253755.288  &               &   \\
45 - 44	&   259390.744  &               &   257443.430\\
46 - 45 &   265025.964  &   255960.830  &   263160.721\\
47 - 46	&               &   261521.591  &   268877.730\\
48 - 47	&               &   267082.092  &   274594.608\\
49 - 48 & & 272642.376  &   \\
50 - 49 & & 278202.381  &  \\
    \bottomrule    
  \end{tabular}
  
  \smallskip
  \begin{minipage}{0.9\textwidth}
  Note: Estimated measurement uncertainties are 25\,kHz. \\
  $^a$ Several of the Lines were previously reported in Refs. \cite{Yamamoto:1987jd,Tang:1995jr}.\\
  $^b$  Centroid used in least-squares fit.\\
  \end{minipage}
  \label{c3s_mm_freqs}
\end{sidewaystable*}

\begin{table*}[t]
    \centering
  \caption{Spectroscopic constants of the ground and vibrationally excited normal and isotopic \ce{C3S} (in MHz, sorted by increasing $B$ values or by isotopic species).  }
  \begin{tabular}{ll l l l l lr r  }
    \toprule
Iso. Species & Vib. State &   $B_v$ &	$10^3D_v$ & $10^9H_v$ &$ q/2$ &$ 10^6q_J/2$  & Weighted ave.$^a$   &  $N^b$  \\
    \midrule
%\cline{1}	          
$^{13}$CCCS		& 	GS   	        & 	2783.06176(6)   & 0.20782(3)   & 0.063 &     &            &    0.62  & 11  \\
$^{13}$CCCS		& 	$v_3=1$	        &   2779.4698(3)   & 0.211(5)   & 0.063 &              &   &    1.40  & 5  \\
C$_3^{\,34}$S  	&	$v_3=1$  	    &   2816.6510(1)   & 0.22441     & 0.063 &              &&    0.45  & 5  \\
C$_3^{\,34}$S  	&	GS      	    & 	2820.36928(6)   & 0.21389(2)   & 0.063 &             &    &    1.05  & 20  \\  
C$_3^{\,34}$S  	&	$v_4=1^1$  	    & 	2825.8842(1)   & 0.212(2)   & 0.063 &    0.71829(5)  & &    0.64  & 10 \\
C$_3^{\,33}$S  	&	GS$^d$      	& 	2854.3868(2)   & 0.222(4)   & 0.063 &             &    &    1.12  & 9  \\
C$^{13}$CCS		& 	$v_3=1$	        & 	2857.5849(1)   & 0.22441     & 0.063 &           &  &    0.49  & 5  \\
C$^{13}$CCS  	&	GS           	& 	2861.37104(6)   & 0.21959(3)   & 0.063 &     &            &    0.98  & 9 \\	
C$_3$S			&   $(v_1,v_3)=(1,2)$  & 	2867.6709(1)   & 0.22441    & 0.063 &             &    &    0.30  & 5  \\
C$_3$S			&   $(v_1,v_3)=(1,1)$  & 	2871.7479(1)   & 0.22441     & 0.063 &           &      &    0.61  & 4  \\
C$_3$S			&   $v_3=4$         &	2875.2412(1)   & 0.22441    & 0.063 &              &  &    0.83  & 4  \\
C$_3$S			&   $v_1=1$         &	2875.5673(1)   & 0.22441     & 0.063 &              &   &    0.23  & 5  \\
C$_3$S			&   $(v_2,v_3)=(1,1)$  &	2875.3676(1)   & 0.22441    & 0.063 &            &     &    0.18  & 5  \\
C$_3$S			&   $v_3=3$         &	2878.9658(1)   & 0.22441    & 0.063 &              &   &    0.41  & 5  \\
C$_3$S			&   $v_2=1$         &	2879.1898(1)   & 0.22441    & 0.063 &             &    &    0.36  & 4  \\
C$_3$S			&   $v_3=2$         &	2882.7415(1)   & 0.22392(5) & 0.063 &       &   &    0.46  & 8  \\
C$_3$S          &   $(v_2,v_4)=(1,1^1)$  &   2884.7028(1)   & 0.22441    & 0.063 &  0.7530(1)  &    &    1.46  & 10  \\
C$_3$S			&   $v_3=1$         &   2886.5512(1)   & 0.22387(4) & 0.063 &       &   &    1.06  & 11 \\				
C$_3$S          &   GS              &   2890.38018(5)  & 0.22441(2) & 0.063(4) &     &            & 0.93 & 41 \\		
C$_3$S			&   $v_4=1^1$	        &	2896.02580(5)  & 0.22756(2)     & 0.079(3) &  0.75353(3) &-0.238(5)    &   0.93  & 92 \\
C$_3$S          &   $(v_3,v_4)=(1,1^1)$  &   2892.92123(7)  & 0.2317(6)  & 2.6(2) &  0.82607(7) &0.41(4)    &    0.77  & 14 \\ 
C$_3$S			&   $v_5=1^1$         &  	2902.70573(7)  & 0.24674(3)   & 0.138(5) &  1.98217(5) &-5.735(9)    &    0.91  & 88 \\
C$_3$S			&   $v_4=2^0$       &   2900.7860(1)   & 0.22007(4) & 0.063 &         & &    0.10  & 10 \\
C$_3$S			&   $v_4=2^2$       &   2901.6282(1)   & 0.23533(4) & 0.063 &         & &    0.38  & 8  \\
%C$_3$S			&$\nu_4+\nu_5^0$&    \\
%C$_3$S			&$\nu_4+\nu_5^2$&      \\
C$_3$S			&   $v_5=2^{0\,c}$       &  2915.1410(1)    & 0.4398(1)    & 1.40(5) &       &          &    0.71 & 43 \\
    \bottomrule
%       \end{tabular}

\\
    \toprule
C$_3$S          &   GS              &   2890.38018(5)  & 0.22441(2) & 0.063(4) &     &            & 0.93 & 41 \\ 		
    			&   $v_1=1$         &	2875.5673(1)   & 0.22441     & 0.063 &              &   &    0.23  & 5  \\
    			&   $v_2=1$         &	2879.1898(1)   & 0.22441    & 0.063 &             &    &    0.36  & 4  \\
    			&   $v_3=1$         &   2886.5512(1)   & 0.22387(4) & 0.063 &       &   &    1.06  & 11 \\			
    			&   $v_3=2$         &	2882.7415(1)   & 0.22392(5) & 0.063 &       &   &    0.46  & 8  \\
    			&   $v_3=3$         &	2878.9658(1)   & 0.22441    & 0.063 &              &   &    0.41  & 5  \\
    			&   $v_3=4$         &	2875.2412(1)   & 0.22441    & 0.063 &              &  &    0.83  & 4  \\
    			&   $v_4=1^1$	        &	2896.02580(5)  & 0.22756(2)     & 0.079(3) &  0.75353(3) &-0.238(5)    &   0.93  & 92 \\
    			&   $v_4=2^0$       &   2900.7860(1)   & 0.22007(4) & 0.063 &         & &    0.10  & 10 \\
    			&   $v_4=2^2$       &   2901.6282(1)   & 0.23533(4) & 0.063 &         & &    0.38  & 8  \\
    			&   $v_5=1^1$         &  	2902.70573(7)  & 0.24674(3)   & 0.138(5) &  1.98217(5) &-5.735(9)    &    0.91  & 88 \\
    			&   $v_5=2^{0\,c}$       &  2915.1410(1)    & 0.4398(1)    & 1.40(5) &       &          &    0.71 & 43 \\
    			&   $(v_1,v_3)=(1,1)$  & 	2871.7479(1)   & 0.22441     & 0.063 &           &      &    0.61  & 4  \\
    			&   $(v_1,v_3)=(1,2)$  & 	2867.6709(1)   & 0.22441    & 0.063 &             &    &    0.30  & 5  \\
    			&   $(v_2,v_3)=(1,1)$  &	2875.3676(1)   & 0.22441    & 0.063 &            &     &    0.18  & 5  \\
              &   $(v_2,v_4)=(1,1^1)$  &   2884.7028(1)   & 0.22441    & 0.063 &  0.7530(1)  &    &    1.46  & 10  \\	
              &   $(v_3,v_4)=(1,1^1)$  &   2892.92123(7)  & 0.2317(6)  & 2.6(2) &  0.82607(7) &0.41(4)    &    0.77  & 14 \\ 

%\cline{1}
C$_3^{\,34}$S  	&	GS      	    & 	2820.36928(6)   & 0.21389(2)   & 0.063 &             &    &    1.05  & 20  \\ 
              	&	$v_3=1$  	    &   2816.6510(1)   & 0.22441     & 0.063 &              &&    0.45  & 5  \\
              	&	$v_4=1^1$  	    & 	2825.8842(1)   & 0.212(2)   & 0.063 &    0.71829(5)  & &    0.64  & 10 \\	          
$^{13}$CCCS		& 	GS   	        & 	2783.06176(6)   & 0.20782(3)   & 0.063 &     &            &    0.62  & 11  \\
        		& 	$v_3=1$	        &   2779.4698(3)   & 0.211(5)   & 0.063 &              &   &    1.40  & 5  \\
C$^{13}$CCS  	&	GS           	& 	2861.37104(6)   & 0.21959(3)   & 0.063 &     &            &    0.98  & 9 \\ 	
        		& 	$v_3=1$	        & 	2857.5849(1)   & 0.22441     & 0.063 &           &  &    0.49  & 5  \\
C$_3^{\,33}$S  	&	GS$^d$      	& 	2854.3868(2)   & 0.222(4)   & 0.063 &             &    &    1.12  & 9  \\ 
        		
    \bottomrule
       \end{tabular}
       
  \smallskip
  \begin{minipage}{\textwidth} \small
        Note: Uncertainties (1$\sigma$) are in units of the last significant digit. Best-fit constants derived from line frequencies in Tables~\ref{c3s_cm_freqs} \& \ref{c3s_mm_freqs} and available pure rotational data from the literature\cite{Ohshima:1992fa, sakai:9831, Yamamoto:1987jd, Tang:1995jr, Crabtree:2016fj}, using a standard linear molecule $^1\Sigma$ Hamiltonian, either with or without $l$-type doubling. Values with no associated uncertainties were constrained to the value of the normal isotopic species.\\
        $^a$ Dimensionless.\\
        $^b$ Refers to the number of lines in the fit.\\
        $^c$ An additional CD-term, $L= 0.207(6) \times 10^{-12}$ MHz, was required to fit the dataset to experimental accuracy.\\
        $^d$ Hyperfine constant: $\chi_{aa}$(S) = -15.889(9) MHz.     
   \end{minipage}
  \label{c3s_constants}
\end{table*}

\begin{table*}[ht]
\centering
  \caption{Measured centimeter-wave transitions of C$_3^{\,33}$S (in MHz).}
  \begin{tabular}{c c}
    \toprule
    $J'_{N'} - J''_{N''}$	&	Frequency	\\
    \midrule	
    $2_2 - 1_1$                           &   11414.8913\\
    $2_4 - 1_3$                           &   11417.7690\\
    $2_2 - 1_2$                           &   11419.6594\\
    $3_3 -  2_2$                           &   17125.7650\\
    $3_4 -  2_3$                           &   17126.4239\\
    $3_3 -  2_3$                           &   17127.6553\\
    $4_6 -  3_5$                           &   22835.1214\\
    $5_7 -  4_6$                           &   28543.8155\\
    $6_8 -  5_7$                           &   34252.4894\\
    \bottomrule
  \end{tabular}
  
  \smallskip
  \begin{minipage}{0.3\textwidth}
   Note:  Estimated measurement uncertainties are 2\,kHz.  
  \end{minipage}
  \label{ccc33s_freqs}
\end{table*}

%%%%%%%%%%%%%%%%%%%%%%%%%%%%%%%%%%%%%%%%%%%%%%%%%%%%%%%%%%%%%%%%%%%%%%%%%%%%%%%%%%%%%%%%%
%%%%%%%%%%%%%%%%%%%%%%%%%%%  C4S data     %%%%%%%%%%%%%%%%%%%%%%%%%%%%%%%%%%
%%%%%%%%%%%%%%%%%%%%%%%%%%%%%%%%%%%%%%%%%%%%%%%%%%%%%%%%%%%%%%%%%%%%%%%%%%%%%%%%%%%%%%%%%

\begin{figure*}[ht]
    \centering
    \includegraphics{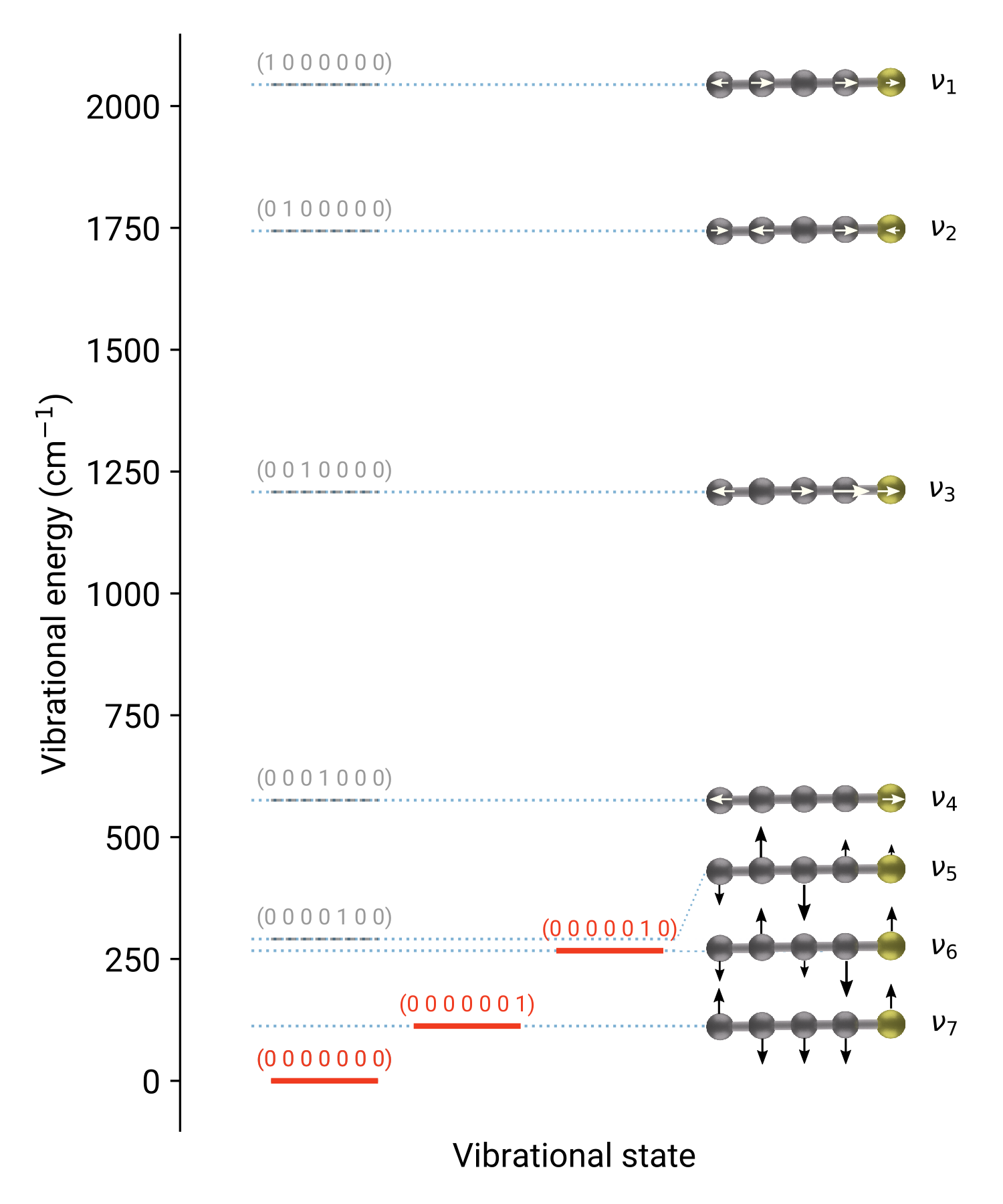}
    \caption{Vibrational states of \ce{C4S}. Due to the high density of states for \ce{C4S}, only observed and fundamental vibrational states are shown. States observed in the experiment are marked as red lines, and unobserved fundamentals are shown as grey dashed lines. The notation for the vibrational quantum numbers is the same as \ce{C3S} in Figure \ref{egy_diag}.}
    \label{egy_diag_c4s}
\end{figure*}

\begin{table*}
\centering
  \caption{Measured centimeter-wave transitions of vibrationally excited \ce{C4S} (in MHz).}
    \begin{tabular}{c r r c r r r r }
    \toprule
     Transition        &  \multicolumn{5}{c}{\ce{C4S}} & & \multicolumn{1}{c}{C$_4^{\,34}$S}\\      
\cmidrule{2-6} \cmidrule{8-8}
$J'_{N'} - J''_{N''}$  &     \multicolumn{2}{c}{$v_6=1^1$}   & &   \multicolumn{2}{c}{$v_7=1^1$} & & \multicolumn{1}{c}{GS} \\
      
    \midrule
  $4_3 -3_2$                     &       8883.3601   &       8885.7015  & &    8884.9911   &    8888.9980 & & 8653.3629 \\
  $5_4 - 4_3$                    &      11845.3056   &      11848.4301  & &   11847.4849   &   11852.8186 & & 11538.5638 \\
  $4_5 -5_4$                     &      14807.9441   &      14811.8526  & &   14810.6815   &   14817.3398 & & 14424.4002 \\
  $5_6 - 4_5$                    &      17771.4438   &      17776.1362  & &   17774.7441   &   17782.7184 & & 17311.0191 \\
                                             
    \bottomrule
     \end{tabular}
     
    \smallskip
    \begin{minipage}{0.8\textwidth}
        Note: Estimated measurement uncertainties are 2\,kHz.\\
    \end{minipage}
  \label{c4s_cm_freqs} 
\end{table*}

\begin{table*}[ht]
\centering
%\small
  \caption{Spectroscopic constants for the ground and newly-identified states of \ce{C4S} (in MHz).}
  \label{c4s_constants}
  \begin{tabular}{@{\extracolsep{\fill}}l D{.}{.}{-1}D{.}{.}{-1} D{.}{.}{-1} D{.}{.}{-1} }
    \toprule
                              & \multicolumn{3}{c}{\ce{C4S}} & \multicolumn{1}{c}{C$_4^{\,34}$S}\\
        \cmidrule{2-4} \cmidrule{5-5}
                		         &	\multicolumn{1}{c}{GS$^a$}	&	\multicolumn{1}{c}{$v_6=1^1$}	& \multicolumn{1}{c}{$v_7=1^1$} & \multicolumn{1}{c}{GS} \\
    \midrule	
	$B$								&    1519.2063(3)		&	 1521.813(9)	&	1522.286(8)   &	1481.2867(1)\\
	$10^6D$					        &	   49.(3)			&    49.			&	49.	& 49.	\\
	$\gamma$						&	   -4.4(7)			&	-4.4 		    &	-4.4	&	-4.4 \\
	$\lambda$						&  113866.(50)			&	114388.(20)		&	114300.(20)	& 113866.	\\
	$10^3\lambda_D$			        &	   12.(3)			&	12.  			&	12.  	&	  12.  	\\
	$p/2$                           &                       &   0.              &   3.7(4) & \\
	$q/2$                           &                       &   -0.20619(8)     &  -0.25(1) & \\
	\vspace{-0.5em}\\
	N. lines$^b$                    &\multicolumn{1}{c}{23} &  \multicolumn{1}{c}{8} &   \multicolumn{1}{c}{8} & \multicolumn{1}{c}{4} \\
	weighted ave.$^c$				&	0.50				&	0.54				&	0.64 &	0.82 \\
    \bottomrule
  \end{tabular}
  
  \smallskip
  \begin{minipage}{0.7\textwidth}
  Note: Uncertainties (1$\sigma$) are in units of the last significant digit. Best-fit constants derived from line frequencies in Table~\ref{c4s_cm_freqs} using a standard linear molecule $^3\Sigma$ Hamiltonian with or without $l$-type doubling. Values with no associated uncertainties were constrained to that of the normal isotopic species.\\
  $^a$ Constants re-fit from Refs. \citenum{Hirahara:1993ud, Gordon:2001pd}.\\ %Ref.~\citenum{Yamamoto:1990wc}
  $^b$ Number of lines in the fit\\
  $^c$ Dimensionless
  \end{minipage}
\end{table*}

\begin{table*}[ht]
\centering
%\small
  \caption{\ Millimeter-wave measurements of the $J=6-5$ transition for vibrationally excited CS, C$^{34}$S, $^{13}$CS, and C$^{33}$S (in MHz). Frequencies reported here are from this work unless otherwise stated.}
  \label{cs_vibstates}
  \begin{tabular}{r llll}
    \toprule
    $v$	& \multicolumn{1}{c}{CS}	&	\multicolumn{1}{c}{C$^{34}$S}	& \multicolumn{1}{c}{$^{13}$CS} & \multicolumn{1}{c}{C$^{33}$S}\\
    \midrule
    0   & 293912.244$^a$ & 289209.230$^a$ & 277455.405$^a$ & 291485.935$^a$ \\
    1   & 291782.294$^a$ & 287130.151$^b$ & 275502.230$^a$ & 289382.425$^a$ \\
    2   & 289651.693$^a$ & 285050.562$^b$ & 273548.395$^a$ & 287278.061 \\ 
    3   & 287520.016$^b$ & 282970.285$^a$ & 271593.790 & 285173.080 \\
    4   & 285387.768$^b$ & 280889.065$^a$ & 269638.443 & \\
    5   & 283254.462$^b$ & 278806.894 & 267682.290 & \\
    6   & 281120.322$^a$ & 276723.786 & 265725.199 & \\
    7   & 278984.935$^b$ & 274639.650 & 263767.219 & \\
	8	& 276848.318$^b$ & 272554.361 & &   \\
	9	& 274710.554$^b$ & 270467.834 &	&   \\
	10	& 272571.341$^b$ & 268379.932$^c$ &	&   \\
	11	& 270430.652$^b$ &	266290.679 &	&   \\
	12	& 268288.286$^b$ &	264199.836 &	&   \\
	13	& 266144.165$^b$ &	262107.273 &	&   \\
	14	& 263998.197$^b$ &	260012.819 &	&   \\
	15	& 261849.952$^b$ &	257916.294 &	&   \\
	16	& 259699.548$^{b,c}$ & & &	\\
	17	& 257546.593 & & &		\\
	18	& 255390.968 & & & 		\\
    \bottomrule
  \end{tabular}
  
  \smallskip
  \begin{minipage}{0.55\textwidth}
  Note: Estimated measurement uncertainty is 25\,kHz. Values not provided are due to gaps in the millimeter-wave survey. No hyperfine splitting was observed for $^{13}$CS and C$^{33}$S.\\
  $^a$ Frequency from Ref. \citenum{ahrens_pure_1999} as the frequency was not covered in our survey.\\
  $^b$ Frequency from our millimeter-wave survey, however has been previously observed in Ref. \citenum{ahrens_pure_1999}\\
  $^c$ Partially blended, but no effect was seen on the frequency uncertainty.\\
  \end{minipage}
\end{table*}

\begin{table*}
  \caption{Measured centimeter-wave transitions of SH in $v=0$ and $v=1$ (in MHz).}
  \label{sh_cm_freq}
\centering
\begin{tabular}{cccccccc D{.}{.}{-1} D{.}{.}{-1}}
\toprule
$N'$ & $J'$ & $p'$ & $N''$ & $J''$ & $p''$ & $F'$ & $F''$ & \multicolumn{1}{c}{v=0} & \multicolumn{1}{c}{v=1} \\ \midrule
 1   & 0.5  & $-$  & 1     &  0.5  & $+$   & 0    & 1     & 8393.3884       & 8075.0265 \\
     &      &      &       &       &       & 1    & 1     & 8445.2093^a  & 8126.3851 \\
     &      &      &       &       &       & 1    & 0     & 8459.0332^a  & 8141.4130 \\
     \bottomrule
\end{tabular}
  
  \smallskip
  \begin{minipage}{0.55\textwidth}
   Note:  Estimated measurement uncertainties are 2\,kHz.  \\
   $^a$ Previously reported with similar uncertainties in Ref. \citenum{Meerts:1975kq}
  \end{minipage}
\end{table*}

\end{document}